\documentclass[longauth]{aa}

\usepackage{amsmath}
\usepackage{graphicx}
\usepackage{graphics}
\usepackage{epsfig}
\usepackage{pdfpages} 
\usepackage{blindtext}
\usepackage{deluxetable}
\usepackage{xspace}

\newcommand{\sii}{\ion{S}{ii}}
\newcommand{\hi}{\ion{H}{i}}
\newcommand{\hii}{\ion{H}{ii}}

\newcommand{\oi}{\ion{O}{i}}
\newcommand{\cii}{\ion{C}{ii}}
\newcommand{\lya}{Ly$\alpha$\xspace}
\newcommand{\oiii}{\ion{O}{iii}}
\newcommand{\oii}{\ion{O}{ii}}
\newcommand{\neiii}{\ion{Ne}{iii}}
\newcommand{\hei}{\ion{He}{i}}
\newcommand{\nii}{\ion{N}{ii}}
\newcommand{\dlya}{$D_{\rm Ly\alpha}$\xspace}

\newcommand{\jw}{\emph{JWST}\xspace}

\usepackage[normalem]{ulem} 
\definecolor{darkgreen}{rgb}{0.0,0.75,0.0}

\usepackage{txfonts}
\usepackage[colorlinks=true,linkcolor=blue,citecolor=blue,urlcolor=blue]{hyperref}%
\begin{document}

\title{The JWST-PRIMAL Legacy Survey}
\subtitle{A JWST/NIRSpec reference sample for the physical properties and Lyman-$\alpha$ absorption and emission of $\sim 500$ galaxies at $z=5.5-13.4$}
\titlerunning{JWST-PRIMAL. A JWST/NIRSpec legacy sample of galaxies at $z=5.5-13.4$}

\author{
K.~E.~Heintz,\inst{1,2,3}
G.~B.~Brammer,\inst{1,2}
D.~Watson,\inst{1,2}
P.~A.~Oesch,\inst{3,1,2}
L.~C.~Keating,\inst{4}
M.~J.~Hayes,\inst{5} \\
Abdurro'uf,\inst{6,7}
K.~Z.~Arellano-C\'ordova,\inst{4}
A.~C.~Carnall,\inst{4}
C.~R.~Christiansen,\inst{1,2}
F.~Cullen,\inst{4}
R.~Davé,\inst{4}
P.~Dayal,\inst{8}
A.~Ferrara,\inst{9}
K.~Finlator,\inst{10,1,2}
J.~P.~U.~Fynbo,\inst{1,2}
S.~R.~Flury,\inst{11}
V.~Gelli,\inst{1,2}
S.~Gillman,\inst{1,12}
R.~Gottumukkala,\inst{1,2}
K.~Gould,\inst{1,2}
T.~R.~Greve,\inst{1,12}
S.~E.~Hardin,\inst{13}
T.~Y.-Y Hsiao,\inst{6,7}
A.~Hutter,\inst{1,2}
P.~Jakobsson,\inst{14}
M.~Killi,\inst{15}
N.~Khosravaninezhad,\inst{16,1,2}
P.~Laursen,\inst{1,2}
M.~M.~Lee,\inst{1,12}
G.~E.~Magdis,\inst{1,12,2}
J.~Matthee,\inst{17}
R.~P.~Naidu,\inst{18}\thanks{NASA Hubble Fellow}
D.~Narayanan,\inst{19,1,2}
C.~Pollock,\inst{4}
M.~Prescott,\inst{10}
V.~Rusakov,\inst{1,2}
M.~Shuntov,\inst{1,2}
A.~Sneppen,\inst{1,2}
R.~Smit,\inst{20}
N.~R.~Tanvir,\inst{21}
C.~Terp,\inst{1,2}
S.~Toft,\inst{1,2}
F.~Valentino,\inst{22,1}
A.~P.~Vijayan,\inst{1,12}
J.~R.~Weaver,\inst{11}
J.~H.~Wise,\inst{13}
J.~Witstok,\inst{23,24}
}

\institute{
Cosmic Dawn Center (DAWN), Denmark 
\and
Niels Bohr Institute, University of Copenhagen, Jagtvej 128, 2200 Copenhagen N, Denmark 
\and
Department of Astronomy, University of Geneva, Chemin Pegasi 51, 1290 Versoix, Switzerland 
\and
Institute for Astronomy, University of Edinburgh, Royal Observatory, Edinburgh EH9 3HJ, UK 
\and
Stockholm University, Department of Astronomy and Oskar Klein Centre for Cosmoparticle Physics, AlbaNova University Centre, SE-10691, Stockholm, Sweden 
\and
Department of Physics and Astronomy, The Johns Hopkins University, 3400 N Charles St. Baltimore, MD 21218, USA 
\and
Space Telescope Science Institute (STScI), 3700 San Martin Drive, Baltimore, MD 21218, USA 
\and
Kapteyn Astronomical Institute, University of Groningen, 9700 AV Groningen, The Netherlands 
\and
Scuola Normale Superiore, Piazza dei Cavalieri 7, 56126 Pisa, Italy  
\and
New Mexico State University, Las Cruces, 88003 NM, USA 
\and
Department of Astronomy, University of Massachusetts Amherst, Amherst, MA 01002, United States 
\and
DTU Space, Technical University of Denmark, Elektrovej, Building 328, 2800, Kgs. Lyngby, Denmark 
\and
Center for Relativistic Astrophysics, School of Physics, Georgia Institute of Technology, 837 State Street, Atlanta, GA 30332, USA 
\and
Centre for Astrophysics and Cosmology, Science Institute, University of Iceland, Dunhagi 5, 107, Reykjavik, Iceland 
\and
Instituto de Estudios Astrofísicos, Facultad de Ingeniería y Ciencias, Universidad Diego Portales, Av. Ejército 441, Santiago 8370191, Chile 
\and
Department of Physics and Astronomy, University of California, Riverside, 900 University Avenue, Riverside, CA 92521, USA 
\and
Institute of Science and Technology Austria (ISTA), Am Campus 1, 3400 Klosterneuburg, Austria 
\and
MIT Kavli Institute for Astrophysics and Space Research, 77 Massachusetts Ave., Cambridge, MA 02139, USA 
\and
Department of Astronomy, University of Florida, 211 Bryant Space Sciences Center, Gainesville, FL 32611, USA 
\and
Astrophysics Research Institute, Liverpool John Moores University, Liverpool, L35 UG, UK 
\and
School of Physics and Astronomy, University of Leicester, University Road, Leicester LE1 7RH, UK 
\and 
European Southern Observatory, Karl-Schwarzschild-Str. 2, 85748 Garching, Germany 
\and
Kavli Institute for Cosmology, University of Cambridge, Madingley Road, Cambridge CB3 0HA, UK 
\and
Cavendish Laboratory, University of Cambridge, 19 JJ Thomson Avenue, Cambridge CB3 0HE, UK 
}
\authorrunning{Heintz et al.}

\date{Submitted ---, accepted ---, published ---}
\abstract{One of the surprising early findings with \jw has been the discovery of a strong ``roll-over'' or a softening of the absorption edge of \lya in a large number of galaxies at \(z\gtrsim 6\), in addition to systematic offsets from photometric redshift estimates and fundamental galaxy scaling relations.
This has been interpreted as strong cumulative damped \lya\ absorption (DLA) wings from high column densities of neutral atomic hydrogen (\hi), signifying major gas accretion events in the formation of these galaxies.}
{To explore this new phenomenon systematically, we assemble the \jw/NIRSpec PRImordial gas Mass AssembLy (PRIMAL) legacy survey of 494 galaxies at $z=5.5-13.4$, designed to study the physical properties and gas in and around galaxies during the reionization epoch. }
{We characterize this benchmark sample in full and spectroscopically derive the galaxy redshifts, metallicities, star-formation rates, and ultraviolet slopes. We define a new diagnostic, the \lya\ damping parameter \dlya\ to measure and quantify the net effect of \lya emission strength, \hi\ fraction in the IGM, or local \hi\ column density for each source. The \jw-PRIMAL survey is based on the spectroscopic DAWN \jw Archive (DJA-Spec).
We describe DJA-Spec in this paper, detailing the reduction methods, the post-processing steps, and basic analysis tools. All the software, reduced spectra, and spectroscopically derived quantities and catalogs are made publicly available in dedicated repositories.}
{We find that the fraction of galaxies showing strong integrated DLAs with $N_{\rm HI} > 10^{21}\,$cm$^{-2}$ only increases slightly from $\approx 60\%$ at $z\approx 6$ up to $\approx 65-90\%$ at $z>8$. Similarly, the prevalence and prominence of \lya\ emission is found to increase with decreasing redshift, in qualitative agreement with previous observational results. Strong \lya\ emitters (LAEs) are predominantly found to be associated with low-metallicity and UV faint galaxies. By contrast, strong DLAs are observed in galaxies with a variety of intrinsic physical properties, but predominantly at high redshifts and low metallicities. }
{Our results indicate that strong DLAs likely reflect a particular early assembly phase of reionization-era galaxies, at which point they are largely dominated by pristine \hi\ gas accretion. At $z=8-10$, this gas gradually cools and forms into stars that ionize their local surroundings, forming large ionized bubbles and produce strong observed \lya\ emission at $z<8$.}

\maketitle

\section{Introduction} \label{sec:intro}

The first epoch of galaxy formation is believed to have occurred at $z\sim 15-20$ \citep{Hashimoto18,Robertson22}, some $100-200$ million years after the Big Bang. This process was primarily driven by the accretion and gravitational collapse of primordial, neutral gas onto dark matter halos \citep{White78,Blumenthal84,White91,Keres05,Schaye10,Dayal18}, which eventually led to the formation of stars in galaxies. The strong UV radiation originating from the first stars and supermassive black holes gradually ionized their immediate and then large-scale surroundings, initiating the reionization epoch, currently estimated to have ended by $z\approx 5-6$ \citep[e.g.][]{Stark16,Dayal18, Keating20,Robertson22,Bosman22,Fan23}. When these first stars ended their lives as supernovae, the pristine gas was chemically enriched with elements formed through explosive nucleosynthesis \citep{Hoyle1954,Cameron1957,Woosley95}. Charting these independent components, their interplay through the ``baryon cycle'' \citep{Tumlinson17,Peroux20}, and their evolution with cosmic time is at the heart of contemporary galaxy formation and evolution studies. 

The launch of the \emph{James Webb Space Telescope} (\jw) has now enabled us to peer deep into this early cosmic epoch, probing the rest-frame optical and ultraviolet (UV) emission of galaxies, potentially all the way back to redshifts $z\gtrsim 15$ \citep{Castellano22,Naidu22,Harikane23,Atek23a,Donnan23,Bouwens23,Wang23a,Austin23}. This leap in near-infrared (NIR) 
capability now enables detailed characterization of the physical properties and baryonic components of galaxies, reaching the first epoch of galaxy formation, the cosmic dawn.   
This epoch seems to contain a larger population of more luminous ($M_{\rm UV} < -20$) and more massive galaxies than expected from theoretical models \citep{Labbe23,Finkelstein23,Franco23,Harikane23,Adams23,Casey23,Bouwens23,Chemerynska23,McLeod24}. 
Various possibilities have been considered to resolve this discrepancy, including a varying initial mass function \citep[IMF;][though see \citealt{Rasmussen23}]{Haslbauer22,Trinca23}, radiation pressure pushing dust away from star-forming regions \citep{Ferrara23}, bursty star-formation histories \citep{Sun23}, or an unexpected overabundance of active galactic nuclei \citep[AGN;][]{Inayoshi22,Pacucci22,dayal2024}. The apparent discrepancy may be exaggerated by selection biases towards the youngest, most highly star-forming galaxies with more bursty star formation histories at $z>10$ \citep{Mason23,Mirocha23,Shen23}, making the galaxies appear temporarily brighter in the UV. As neutral gas accretion is the primary driver and fuel for star formation, a robust measure of this component is key to resolving the debate. 

Recent \jw spectroscopic studies have pushed the limit of the most distant galaxies confirmed to $z\approx 11$--13 \citep{CurtisLake23,Wang23b,Fujimoto23_uncover,Zavala24,Castellano24}.
The metal content of galaxies during the reionization epoch can now also be readily measured based on direct $T_e$-based methods for sufficiently bright galaxies \citep[e.g.,][]{Schaerer22,ArellanoCordova22,Taylor22,Brinchmann23,Curti23a,Katz23,Rhoads23,Trump23,Heintz23_JWSTALMA,Nakajima23,Sanders24} or inferred through rest-frame optical strong-line calibrations \citep{Heintz23_FMR,Langeroodi23,Curti23b,Matthee23}. These observations show a substantial offset from the fundamental-metallicity relation \citep[FMR,][]{Mannucci2010,Lara-Lopez2010,Curti20,Sanders21} established at $z\approx 0-3$ by up to 0.5\,dex at $z>7$ \citep{Heintz23_FMR}, suggesting copious pristine \hi\ gas inflows onto galaxies at this epoch before chemical enrichment from star formation can catch up to equilibrium \citep[e.g.,][]{Torrey18}. The exact redshift for this transition is still debated \citep{Curti23b,Nakajima23}. 

Constraining the inflow of neutral, pristine \hi\ gas is key to understanding the assembly of the primordial matter and the build-up of stars and metals in the first galaxies. However, the circumstantial evidence from the FMR for excessive \hi\ gas accretion needs to be corroborated by measuring directly the neutral, atomic hydrogen (\hi) -- the key missing element in the baryonic matter budget of high-redshift galaxies.  
Due to the weakness of the hyperfine \hi\ 21\,cm line, \hi\ has historically been measured at redshifts $z\gtrsim 2$ through Lyman-$\alpha$ (\lya) absorption in bright background sources such as quasars \citep{Wolfe86,Wolfe05,Prochaska09,Noterdaeme12,Peroux20} or gamma-ray bursts \citep{Jakobsson06,Prochaska07,Fynbo09,Tanvir19,Heintz23_GRB}. However, these only probe \hi\ in narrow, pencil-beam sightlines and often at large impact parameters. \lya absorption has also been observed directly in integrated spectra of Lyman-break galaxies at $z\approx 3$ \citep{Pettini00,Shapley03,Steidel10,Cooke15,Lin23}, but only in rare, extreme cases. 

Recently, \citet{Heintz23_DLA} reported the  discovery of extremely strong ($N_{\rm HI}\gtrsim 10^{22}$\,cm$^{-2}$) damped \lya absorption systems (DLAs) in star-forming galaxies at $z>8$ \citep[see also][]{Umeda23,DEugenio23,ChenMason23}. These seem to represent a galaxy population showing more prominent and substantially more prevalent DLAs than observed in Lyman-break galaxies at $z\approx 3$ \citep[e.g.][]{Shapley03}. The shape of the \lya damping wings of galaxies had until these discoveries mainly been thought to trace the bulk \hi\ of the intergalactic medium (IGM) at $z\gtrsim 6$ \citep{MiraldaEscude98,McQuinn08}, and had been used to infer the reionization history of the Universe with \jw spectroscopy \citep{Chen23,Keating23a,Keating23b,Umeda23}. The discovery of these massive interstellar or circumgalactic \hi\ gas reservoirs carry important implications for our study of the early Universe: they constitute the first direct evidence for the mass accretion of primordial intergalactic gas into ordinary galaxies, i.e.\ these DLAs represent the creation of the baryonic component of galaxies. They also strongly affect observations of early galaxies because they change the shape of the \lya damping wings, altering and complicating inferences on the state of the bulk IGM. Further, they hinder the escape of ionizing photons \citep{Laursen11,Verhamme15,Steidel18,Hu23,Hayes23}, and may also be responsible for the bias in photometric redshifts compared to spectroscopic measurements \citep{Heintz23_DLA,Finkelstein23b,Fujimoto23_ceers,Hainline23} of galaxies during the reionization era. Their discovery, however, also presents a fortuitous new avenue to directly probe the build-up of pristine \hi\ in galaxies at this critical epoch. 

In this work, we systematically characterize these effects by measuring the prevalence and prominence of strong \lya emission and damped \lya absorption in a large spectroscopic sample of star-forming galaxies at $z=5.5$--\,13.4 observed with \jw/NIRSpec. The observational data for these galaxies are all presented here as part of the DAWN \jw Archive (DJA), which includes fully reduced and post-processed 1D and 2D spectra of all public \jw/NIRSpec data. The large \jw/NIRSpec archive forms the basis of the \jw/NIRSpec PRImordial gas Mass AssembLy (PRIMAL) legacy survey presented here. In this first part of the survey we aim to disentangle the \lya damping wings produced from \hi\ gas in the immediate surroundings of galaxies (ISM or CGM) from the effects of an increasingly neutral IGM, and establish statistical correlations for these features and the presence of \lya emission with the physical properties of the galaxies. We further make the reduced spectra, source catalogs, spectroscopic redshifts, line identifications and fluxes, and the physical properties of each of the galaxies in the \jw-PRIMAL sample publicly available on dedicated webpages. 

\begin{table*}[!th]
\caption{Archival references for the \jw-PRIMAL sample.} 
\label{tab:obs}      
\setlength\tabcolsep{0.6cm}
\begin{tabular}{c c c c c c c}
\hline\vspace{0.1cm}
R.A. (deg) & Decl. (deg) & Prog. ID & Src. ID & S/N$_{\rm UV}$ & $z_{\rm spec}$ & Ref. \\  
 (1) &  (2) &  (3) &  (4) &  (5) & (6) & (7) \\
\hline    
$53.115717$  &	$-27.774955$ &	1210 &	16374  &	60.53 &	5.5050 & (B23) \\
$64.419078$  &	$-11.905924$ &  1208 &	6521   &	13.39 &	5.5050 & -- \\
$214.81967$  &	$52.879755$  & 	1345 &	381    &	12.02 &	5.5161 & -- \\
$53.145647$  &	$-27.801499$ &  3215 &	201906 &	12.24 &	5.5194 & -- \\
$53.139782$  &  $-27.853438$ &  2198 &  10840  &    30.90   & 5.5207 & -- \\
$\vdots$     &   $\vdots$    & $\vdots$ & $\vdots$ & $\vdots$ & $\vdots$ & $\vdots$ \\
$214.90663$ & $52.945504$    & 2750  & 10 &  25.564 & 11.49062 & (AH23) \\
$53.164762$ & $-27.774626$   & 3215  & 20130158 &   61.61 & 11.7060 & (B23, ECL23) \\
$53.166346$ & $-27.821557$   & 3215  & 20096216 &  42.17  & 12.5119 & (ECL23) \\ 
$3.513563$  & $-30.3568$     & 2561  & 38766   &   24.95  & 12.7822 & (W23) \\
$53.149881$ & $-27.776502$   & 3215  & 20128771 &   46.07 & 13.3605 & (ECL23) \\ 
\hline          
\end{tabular} \\
\textbf{Notes.} Column (1): Right ascension in degrees (J\,2000). Column (2): Declination in degrees (J\,2000). Column (3): Program ID under which the object was observed. Column (4): Designated MSA source ID. Column (5): Rest-frame UV signal-to-noise ratio. Column (6): Spectroscopic redshift. Column (7): The original survey paper references. A full version of this table can be found online. \\
{\bf References.} (B23)~\citet{Bunker23_jades}; (AH23)~\citet{ArrabalHaro23}; (ECL23)~\citet{CurtisLake23}; (W23)~\citet{Wang23b}.
\end{table*}

We have structured the paper as follows. In Sect.~\ref{sec:obs} we present the observations, describe the spectroscopic reduction and post-processing, and outline the \jw-PRIMAL sample compilation. In Sect.~\ref{sec:res}, we detail the spectroscopic analysis of \lya, the nebular emission line fluxes and equivalent widths, and the inferred physical properties of the sample galaxies. In Sect.~\ref{sec:dlas}, we consider the cosmic evolution of galaxy DLAs and \lya\ emitters and quantify the physical correlations driving or hampering the observed \lya emission and excess damped \lya absorption. In Sect.~\ref{sec:disc}, we discuss and conclude our work and provide a future outlook. All the spectroscopic data products and the analysis and results presented in this work are made publicly available on dedicated webpages. 
Throughout the paper, we assume concordance flat $\Lambda$CDM cosmology, with $H_0 = 67.4$\,km\,s$^{-1}$\,Mpc$^{-1}$, $\Omega_{\rm m} = 0.315$, and $\Omega_{\Lambda} = 0.685$ \citep{Planck18}. 


\section{Observations, data processing, and sample selection} \label{sec:obs}

\subsection{The DAWN JWST Archive (DJA) -- Spectroscopy}

The data considered in this work are processed as part of the DAWN \jw Archive (DJA), an online repository containing reduced images, photometric catalogs, and spectroscopic data for public \jw data products\footnote{\url{https://dawn-cph.github.io/dja}}. Here, we detail the spectroscopic data reduction and post-processing only, but see \citet{Valentino23} for details on the imaging and photometric data. The DJA spectroscopic archive (DJA-Spec) is, at the time of writing, comprised of observations taken from some of the large Early Release Science (ERS), General Observer (GO) and Guaranteed Time (GTO) Cycle 1 \& 2 programs, such as CEERS \citep[ERS-1345;][]{Finkelstein23}, GLASS-DDT \citep[DDT-2756][]{Treu22,RobertsBorsani23} \citep[ERS-1324;][]{Treu22}, JADES \citep[GTO-1180, 1210, GO-3215;][]{Bunker23_jades,Eisenstein23}, and UNCOVER \citep[GO-2561;][]{Bezanson23}. Many of the spectra analyzed in this work have been presented in dedicated survey papers cited above and in others \citep[e.g.,][]{ArrabalHaro23,Atek23b,Cameron23}, which also describe the sample selection and observational design of the different programs. With DJA-Spec we reduce and extract all of the spectra from this diverse array of observational programs with a single standardized pipeline. 

\subsection{DJA spectroscopic reduction and data processing}


All data processing is performed with the publicly available {\sc Grizli} \citep{Brammer_grizli}\footnote{\url{https://github.com/gbrammer/grizli}} and {\sc MSAExp} \citep{Brammer_msaexp}\footnote{\url{https://github.com/gbrammer/msaexp}} software modules, with the procedure as follows. The spectroscopic analysis begins with reprocessing the individual uncalibrated (\texttt{uncal}) exposures retrieved from the Mikulski Archive for Space Telescopes (MAST) with the \jw\ \href{https://jwst-pipeline.readthedocs.io/en/latest/jwst/pipeline/calwebb_detector1.html#calwebb-detector1}{Detector1Pipeline}.\footnote{{\sc jwst} pipeline version 1.12.5; \texttt{CRDS\_CONTEXT = jwst\_1180.pmap}}. The pipeline is executed with the default parameters, but with \texttt{snowblind}\footnote{\url{https://github.com/mpi-astronomy/snowblind}} run between the \texttt{jump} and \texttt{ramp\_fit} steps for improved masking of the bright cosmic ray ``snowballs''\citep{Rigby23}. We remove a column average from the count-rate (\texttt{rate}) files produced by the first step to remove the $1/f$ noise. We also find that scaling the read noise extension of the calibrated count-rate images by a factor of order $\approx1.4$ derived separately for each exposure is necessary to explain the variance of pixels in un-illuminated portions of the NIRSpec detectors. The corrected \texttt{rate} files are then run through the \href{https://jwst-pipeline.readthedocs.io/en/latest/jwst/pipeline/calwebb_spec2.html}{calwebb\_spec2} pipeline with its default parameters up through the photometric calibration (\texttt{photom}) step. The products at this stage are flux- and wavelength-calibrated 2D spectra for every source indicated with an open shutter in the attached MSA metadata, in a frame cut out from the original detector pixel grid with curved spectral traces. These 2D \href{https://jwst-pipeline.readthedocs.io/en/stable/api/jwst.datamodels.SlitModel.html}{SlitModel} cutouts are saved to individual multi-extension FITS files, and it is here that the subsequent processing with {\sc MSAExp} departs significantly from the standard {\sc jwst} pipeline.

\begin{figure*}[!t]
    \centering
    \includegraphics[width=16cm]{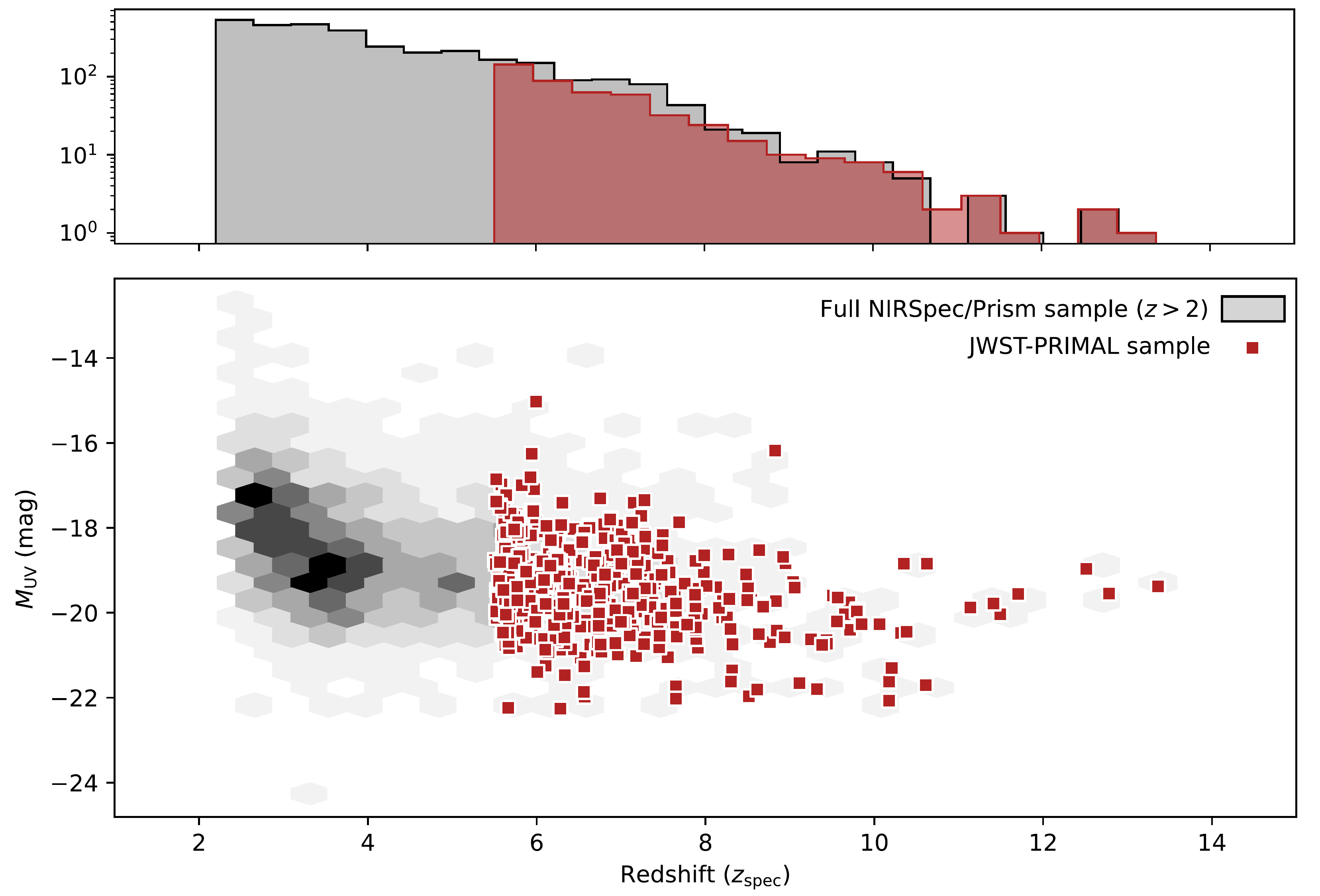}
    \caption{The absolute rest-frame UV magnitude ($M_{\rm UV}$) as a function of spectroscopic redshift for the \jw/NIRSpec Prism sources at $z>2$ from DJA with robust spectroscopic redshifts (grade 3) shown in grey. The \jw-PRIMAL sources are highlighted by the red squares, selected by requiring $z_{\rm spec} > 5.5$, an integrated ${\rm S/N}>3$ around the redshifted \lya\ wavelength region, and full NIRSpec Prism spectroscopic wavelength coverage from 0.6--5.3\ $\mu$m as outlined in Sect.~\ref{ssec:primalsel}.}
    \label{fig:muvzsp}
\end{figure*}

The sky background of each source spectrum is removed by taking differences of the 2D cutout spectra taken at different ``nod'' positions of the telescope. Typically there are three nod offsets that shift the source by $0\farcs5 \approx 5~\mathrm{pixels}$ within the MSA slitlet (the so-called 3-Shutter-Slitlet pattern), though the pipeline calculates the groupings automatically to handle other configurations of the heterogeneous archival observations. Average 2D source and sky spectra (and associated variance images) are calculated from exposures at a given nod position and from those at all other offset positions, respectively. We fit a 2D Gaussian profile to the $source - sky$ spectra where the width is the quadrature sum of a free parameter and the wavelength-dependent width of the point spread function and where a centroid shift is fit starting from an initial guess provided by the exposure MSA metadata and the catalog used to design the MSA mask. We note that the latter centroid shift is usually of order 0.1\,pix (10\,mas), which, while consistent with the expected precision of the input catalog astrometry and telescope pointing, is non-trivial and measurable and required for robust extraction of the 1D spectra. The sky-subtracted 2D spectra cutout spectra at different offset positions are combined in a single rectified frame with orthogonal wavelength and cross-dispersion pixel axes by binning with a 2D histogram algorithm. Finally, the one-dimensional spectra and uncertainties are extracted with ``optimal'' weighting \citep{Horne86} using the fitted 2D Gaussian profile. We note that \citet{RobertsBorsani24} have recently used the same pipeline presented here and developed for this work, presenting a similar effort to spectroscopically characterize galaxies at $z>5$ observed with NIRSpec Prism. 



Currently (Apr 2, 2024) DJA-Spec includes 7,319 individual combined spectra taken with the NIRSpec Prism/CLEAR and 1,665 combined spectra with NIRSpec medium and high-resolution gratings. A full overview can be found on the dedicated webpage\footnote{\url{https://s3.amazonaws.com/msaexp-nirspec/extractions/nirspec_graded_v2.html}}, where the following data are available: best-fit and inspected redshifts, the median signal-to-noise ratios (S/N) of the spectra, grades from visual inspection on the quality of the spectra, \emph{Hubble Space Telescope} (\emph{HST}) and \jw/NIRCam imaging thumbnails, and optimally reduced and extracted 2D and 1D spectra. 

In this work, we focus on the \jw/NIRSpec Prism spectroscopic data from DJA due to the higher continuum sensitivity which is essential for quantifying the evidence for Ly$\alpha$ emission or absorption damping wings. 
\jw's NIRSpec Prism configuration covers the entire near-infrared passband from 0.6\ $\mu$m to 5.3\ $\mu$m, with a varying spectral resolution from $\mathcal{R}\approx 50$ in the blue end to $\mathcal{R}\approx 400$ in the red end \citep{Jakobsen22}. We have carefully reduced and processed the spectra in DJA in a homogeneous way, limiting any biases between reduction and calibration codes across surveys. Further, this optimizes the consistency and uniformity of the \jw-PRIMAL sample measurements and the DJA-Spec sample as a whole.  

\begin{figure*}[!ht]
    \centering
    \includegraphics[width=18cm]{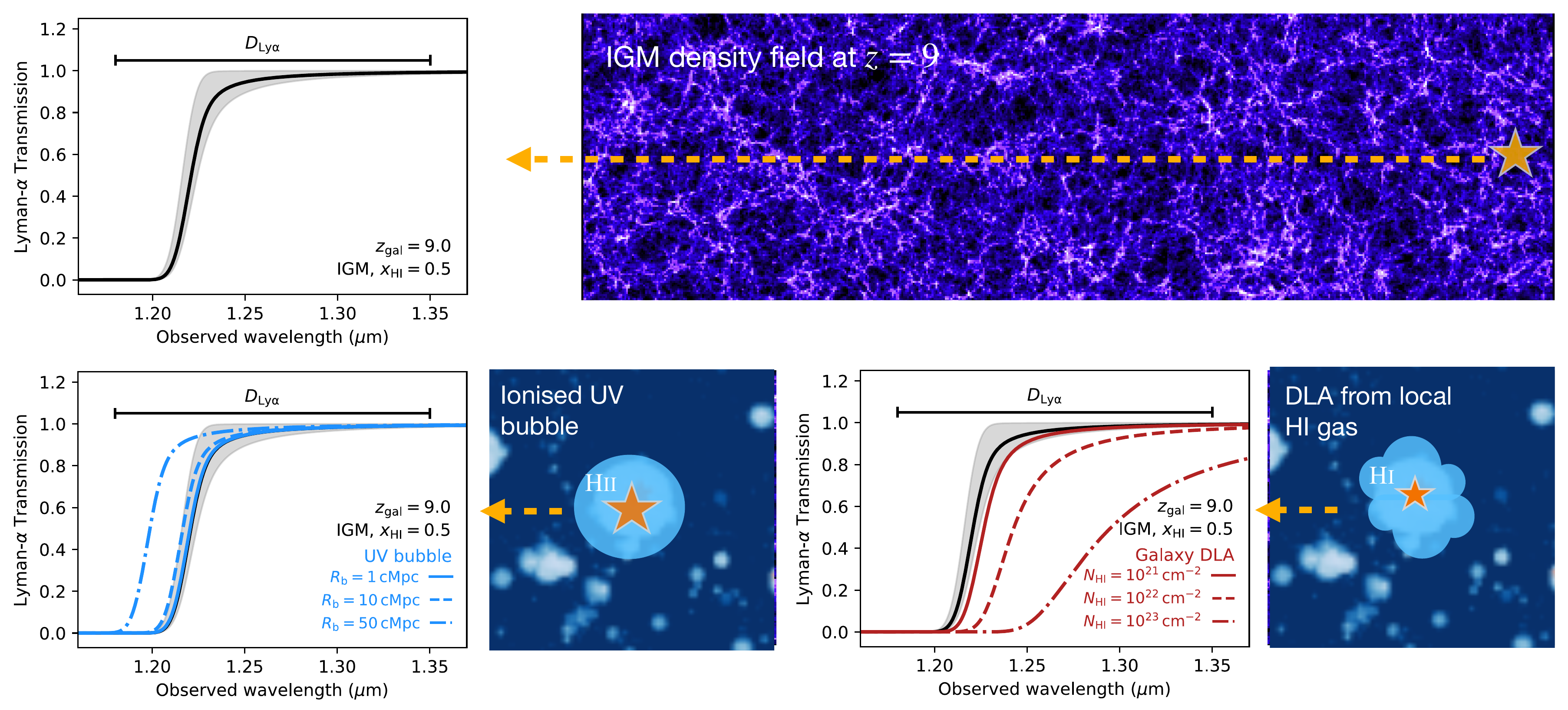}
    \caption{Schematic highlighting the intrinsic physical models affecting the shape of the \lya\ transmission from an example galaxy at $z=9.0$, all convolved by the nominal \jw/NIRSpec Prism spectral resolution. The default model with $x_{\rm HI} = 0.5$ is shown at the top left, where the grey-shaded region represents the effect of a partly ($x_{\rm HI} = 0.01$) to fully neutral ($x_{\rm HI} = 1.0$) IGM. For illustrative purposes the expected IGM density field at $z=9$ is shown in the top right, extracted from the {\sc Astraeus} simulations. In the bottom panels are shown the combined effects of the default IGM model and various sizes of the ionized UV bubble (left, blue: $R_{\rm b} = 1,\,10,\,50$\,cMpc) and DLAs from local \hi\ gas reservoirs (right, red: $N_{\rm HI} = 10^{21},\,10^{22},\,10^{23}$\,cm$^{-2}$). The integration region for $D_{\rm Ly\alpha}$ is marked by the top line in all the \lya\ transmission curve figures and is defined to encapsulate all the predicted physical scenarios.}
    \label{fig:dlyaillu}
\end{figure*}

\subsection{JWST-PRIMAL sample selection} \label{ssec:primalsel}

The main objective of this paper is to accurately model the \lya damping wings and/or measure strong \lya emission, and spectroscopically derive the physical properties of galaxies during the reionization epoch. For this, we require the following criteria for the sources to enter our archival \jw-PRIMAL sample: 
\begin{enumerate}
    \item A robust, spectroscopic redshift measurement at $z>5.5$ from a minimum of one emission line detected at $>3\sigma$ (in addition to the \lya\ break). 
    \item An integrated signal-to-noise over the redshifted \lya\ and rest-frame UV ($1550\,\AA$) regions of ${\rm S/N} > 3$. 
    \item Flux density measurements over the entire NIRSpec Prism wavelength coverage from $0.6-5.3\,\mu$m. 
\end{enumerate}
These criteria ensure that the redshift of the foreground \hi\ gas absorber, from interstellar to intergalactic scales, and the shape of the \lya\ line profile can be accurately measured. Further, the requirement of the full spectral coverage enables a full characterization of each source in the \jw-PRIMAL sample, including robust spectroscopic modelling of the stellar continuum and rest-frame UV spectral slope, emission line fluxes and equivalent widths EWs, and direct constraints on the star-formation rates (SFRs), metallicities, and ionization parameters for all sources. 

In total, {\bf 494} sources from DJA-Spec meet these criteria. Fig.~\ref{fig:muvzsp} show the absolute UV magnitude, $M_{\rm UV}$, as a function of redshift for the \jw-PRIMAL sources and compared to the underlying DJA-Spec sample (at $z>2$). Table~\ref{tab:obs} summarizes the target list, including their coordinates, original program and source ID, and emission-line redshifts. 

\section{Analysis and results} \label{sec:res}

Here we detail the spectroscopic measurements derived for the full \jw-PRIMAL sample. We focus on the \lya\ damping wings, but also detail the measurements and basic physical properties of the sample galaxies. All the spectroscopically-derived quantities are made available on a dedicated webpage\footnote{\url{https://github.com/keheintz/jwst-primal}}.

\subsection{The Lyman-$\alpha$ damping parameter}
\label{ssec:dlya}

During the reionization epoch at $z\gtrsim 6$, several factors add to the line shape of \lya\ on integrated galaxy spectra: They can show strong \lya\ emission \citep[LAE;][]{Matthee18,Mason18,Witstok23}, damping wings from an increasingly neutral IGM \citep{MiraldaEscude98,Keating23a,Chen23}, excess continuum flux in the wings related to the size of the immediate ionized bubbles \citep{McQuinn08,Castellano16,Castellano18,Fujimoto23_uncover,Umeda23,Hayes23}, strong damped \lya\ absorption (DLA) from local \hi\ gas \citep{Heintz23_DLA,DEugenio23}, intrinsic variations due to a changing velocity and density distribution of gas and dust in the ISM and CGM \citep{Dayal11,Verhamme15,Verhamme17,Gronke17,Hutter23} or even possibly be dominated by two-photon emission processes \citep{Steidel16,Chisholm19,Cameron23_2phot}. Many of these effects are degenerate, so we have to disentangle them statistically. For this, we define a new simple diagnostic, which we denote the {\em \lya\ damping parameter}:
\begin{equation}
    D_{\rm Ly\alpha} \equiv \int^{\lambda_{\rm Ly\alpha,up}}_{\lambda_{\rm Ly\alpha,low}} (1-F_\lambda/F_{\rm cont}) ~ {\rm d}\lambda ~ / ~ (1+z_{\rm spec}),
\end{equation}
which is the equivalent width (EW) of the transmitted flux density, $F_\lambda$ over the wavelength region covered by the instrumentally-broadened \lya\ transition. The continuum flux, $F_{\rm cont}$, over the same region is estimated by extrapolating the rest-frame UV slope, $\beta_{\rm UV}$, which is derived directly in each spectra from rest-frame $1400-2600\,\AA$ (see Sect.~\ref{ssec:uv} for further details). This approximation of the stellar continuum is generally consistent with models of the continuum flux using galaxy templates \citep{Heintz23_DLA}. We define the integration limits from $\lambda_{\rm Ly\alpha,low} = 1180\,\AA$ to $\lambda_{\rm Ly\alpha,up}= 1350\,\AA$ (rest-frame) to capture most of the damping feature for the most extreme cases and to simultaneously limit the contamination from absorption or emission lines at longer wavelengths. We tested various lower limits and found this to be the bluest wavelength at which information is not lost at the resolution of NIRSpec Prism. 

A set of physical models, shown as imprints on the \lya\ transmission curves, are visualized in Fig.~\ref{fig:dlyaillu} including: varying ionized bubble sizes from $R_{\rm b} = 1 - 50$\,cMpc, neutral hydrogen fractions of the IGM from $x_{\rm HI} = 0.1-1.0$, and DLA \hi\ column densities $N_{\rm HI} = 10^{21}-10^{22}$\,cm$^{-2}$, all affecting the shape of the \lya\ damping wings. The IGM models are based on the prescription by \citet{MiraldaEscude98}, using the approximation described in \citet{Totani06}, and the UV bubble sizes are included by varying the upper bound of the IGM integration region, such that $z_{\rm IGM,upper} < z_{\rm gal}$ \citep[e.g.,][]{McQuinn08}. The DLAs with varying \hi\ column densities are derived from the Voigt-profile approximation by \citet{TepperGarcia06}. The corresponding \lya\ damping parameter \dlya\ is listed for each model. The UV bubble and DLA models all assume a benchmark IGM fraction of $x_{\rm H\textsc{i}}=0.5$, and may therefore vary with the same scatter in \dlya as observed for the IGM-only models.

\begin{figure}[!t]
    \centering
    \includegraphics[width=9.1cm]{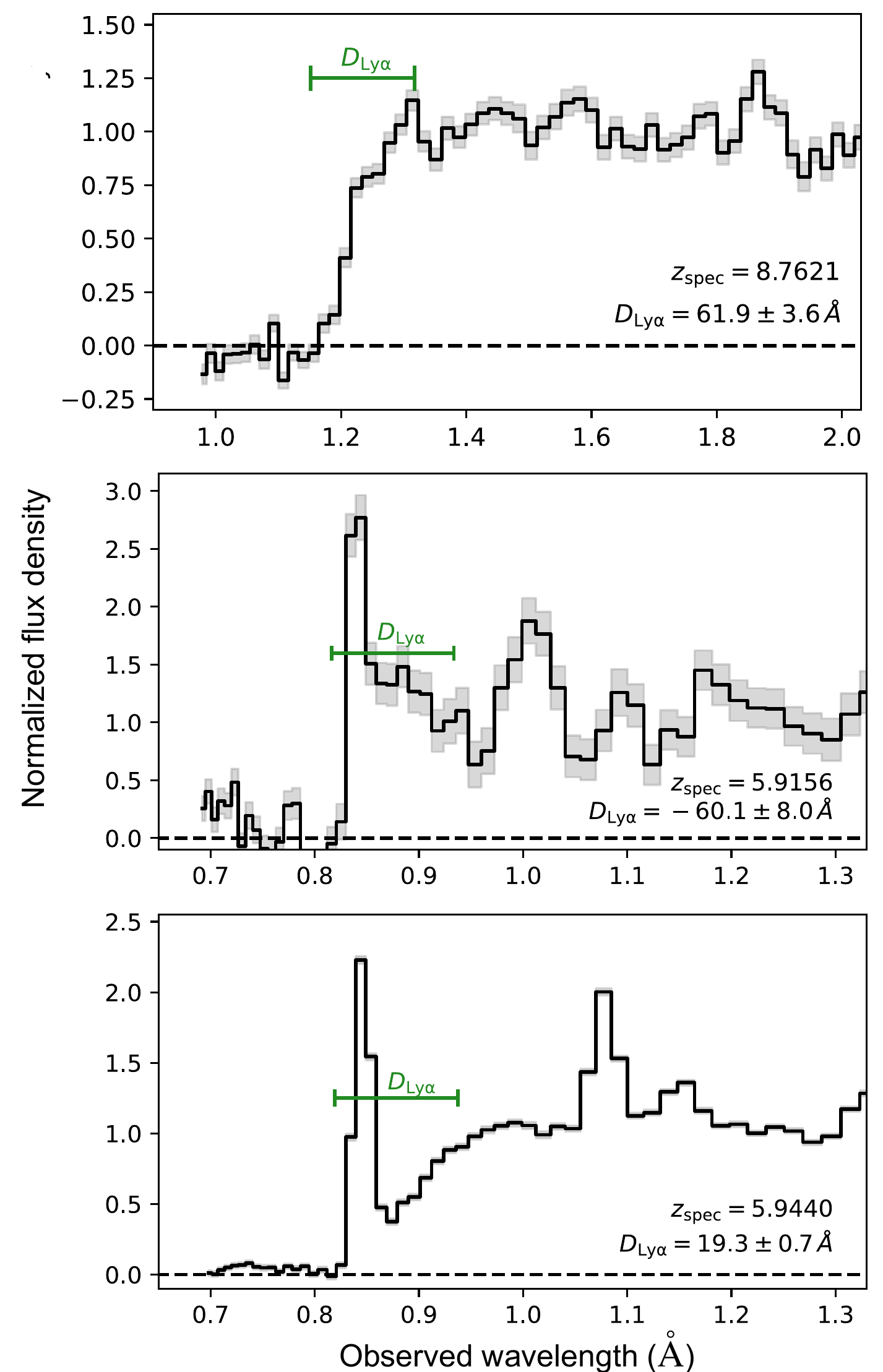}
    \caption{Three examples of normalized NIRSpec Prism spectra from the \jw-PRIMAL sample. The top panel shows an example of a strong DLA from local \hi\ absorption, the middle panel a strong LAE, and the bottom panel a combination of both. The redshifts, \dlya\ measurements, and \dlya\ integration regions are highlighted for each case. }
    \label{fig:dlyaex}
\end{figure}

We derive damping parameters for the full sample in the range $D_{\rm Ly\alpha} = -400$ to $150\,\AA$. With this, we define four parameter spaces probing different physical regimes: (i) cases with excess continuum flux suggesting contributions from ionized bubbles or weak LAEs have $D_{\rm Ly\alpha} = 0$--35\,$\AA$; (ii) the effect on the \lya\ damping wings of an increasing neutral IGM results in $D_{\rm Ly\alpha} = 35$--50\,$\AA$; (iii) excess damped \lya\ absorption from local \hi\ (with $N_{\rm HI}>10^{21}$\,cm$^{-2}$) have $D_{\rm Ly\alpha} > 50\,\AA$; and (iv) strong LAEs are typically observed with $D_{\rm Ly\alpha}<0\,\AA$. Fig.~\ref{fig:dlyaex} shows three example spectra from the \jw-PRIMAL sample. One shows an example of a broad \lya\ damping wing consistent with local \hi\ absorption with $D_{\rm Ly\alpha} = 61.9\pm 3.6\,\AA$, consistent with $N_{\rm HI}\approx 10^{22}\,$cm$^{-2}$. Another shows strong \lya\ emission with a measured $D_{\rm Ly\alpha} = -60.1\pm 8.0\,\AA$. 

This simple diagnostic now allows us to quantify the number of sources in each distinct population.
We caution, however, that the low spectral resolution of the \jw/NIRSpec Prism configuration may conceal weak \lya\ emission in the continuum, which requires higher resolution spectroscopy to resolve \citep[see e.g. the case for GN-z11;][]{Bunker23_gnz11}. This will likely cause a slight decrease in \dlya. Further, as demonstrated in the bottom panel of Fig.~\ref{fig:dlyaex}, rare cases ($\lesssim 1\%$ of the sample) that show both a substantial damped \lya\ broadening and simultaneously a strong \lya\ emission line will not be accurately identified, here with $D_{\rm Ly\alpha} = 19.3\pm 0.7\,\AA$, misleadingly suggesting mild excess \lya\ continuum flux \citep[the continuum flux of this particular source might be dominated by two-photon nebular emission, e.g.,][]{Cameron23_2phot}. The \jw/NIRSpec shutter may also not cover the regions of the strongest \lya\ emission, thereby underestimating the total \lya\ photon output of the target source \citep{Jiang23}. The \dlya\ parameter is thus most powerful in identifying the most extreme cases, such as strong \lya\ emitters or galaxy DLAs with $N_{\rm HI}>10^{21}$\,cm$^{-2}$, which uniquely have $D_{\rm Ly\alpha}<0\,\AA$ and $D_{\rm Ly\alpha}>50\,\AA$, respectively.

\begin{table*}[!t]
\caption{Line flux measurements for the most prominent nebular emission lines \jw-PRIMAL sample.} 
\label{tab:lineflux}      
\begin{tabular}{c c c c c c c c c}
\hline\vspace{0.1cm}
$z_{\rm spec}$ & [\oii]\,$\lambda 3727$ & [\neiii]\,$\lambda 3870$ + & H$\delta$ & H$\gamma$ + & H$\beta$ & [\oiii]\,$\lambda 5008$ & H$\alpha$ + & [\sii]\,6725 \\  
& & \hei\,$\lambda 3889$ & & [\oiii]\,$\lambda 4363$ & & & [\nii]\,$\lambda 6585$ & \\
 (1) &  (2) &  (3) &  (4) &  (5) & (6) & (7) & (8) & (9) \\
\hline   
5.5050 & $54.7\pm 6.1$ & $49.9\pm 7.4$ & $13.8\pm 4.3$ & $21.1\pm 4.9$ & $76.9\pm 3.5$ & $153.5\pm 4.3$ & $252.6\pm 5.0$ & $16.1\pm 3.2$ \\
5.5050 & $<97.5$ & $<150.0$ & $<60.0$ & $175.6\pm 19.8$ & $331.0\pm 20.4$ & $1004.4\pm 31.6$ & $645.5\pm 26.3$ & $< 53.4$ \\
5.5161  & $<132.0$ & $156.0$ & $<100.2$ & $<115.2$ & $84.9\pm 25.7$ & $335.\pm 29.6$ & $359.8\pm 30.9$ & $<78.9$ \\
5.5194 & $<12.9$ & $<16.5$ & $11.4\pm 3.5$	& $12.4\pm 3.5$ & $27.1\pm 2.5$ & $72.4\pm 2.5$ & $84.4\pm 2.8$ & $<8.1$ \\
5.5207 & $219.8\pm 50.1$ & $243.4\pm 54.5$ & $97.4\pm 37.2$ & $<146.7$ & $205.1\pm 33.2$ & $1176.7\pm 46.2$ & $705.8\pm 43.5$ & $<97.5$ \\
$\vdots$ & $\vdots$ & $\vdots$ & $\vdots$ & $\vdots$ & $\vdots$ \\
11.4906 & $17.0\pm 5.0$ & $<36.0,<36.0$ & $<36.0$ & $<45.0$ & $-$ & $-$ & $-$ & $-$ \\
11.7060 & $<8.7$ & $7.3\pm 2.3, <9.6$ & $<11.0$ & $-$ & $-$ & $-$ & $-$ & $-$ \\
12.5119 & $9.1\pm 2.7$ & $9.3\pm 3.1, <11$ & $-$ &  $-$ & $-$ & $-$ & $-$  \\
12.7822 & $<10.0$ & $18\pm 6, <10.0$ & $-$ & $-$ & $-$ & $-$ & $-$ & $-$ \\
13.3606 & $6.0\pm 2.0$ & $-$ & $-$ & $-$ & $-$ & $-$ & $-$ & $-$ \\
\hline          
\end{tabular} 
\textbf{Notes.} All measurements are reported in units of $10^{-20}$\,erg\,s$^{-1}$\,cm$^{-2}$. Uncertainties on measurements are stated at $1\sigma$ standard deviations and upper limits at $2\sigma$ confidence limits. Column (1): Spectroscopic redshift. Column (2): Unresolved [\oii]\,$\lambda 3726,3729$ doublet line flux. Column (3): Unresolved [\neiii]\,$\lambda 3870$ + \hei\,$\lambda 3889$ line flux measurements, except for the highest redshift sources ($z > 10$). Column (4): Balmer H$\delta$. Column (5): Unresolved Balmer H$\gamma$ + auroral [\oiii]\,$\lambda 4363$ line flux measurements, except for the highest redshift sources ($z > 9.5$). Column (6): Balmer H$\beta$. Column (7): [\oiii]\,$\lambda 5008$ line flux, where [\oiii]\,$\lambda 4960$ is assumed to be $1/3$ of this. Column (8): Unresolved Balmer H$\alpha$+[\nii]\,$\lambda 6585$. Column (9): Unresolved [\sii]\,$\lambda\lambda 6718, 6732$ doublet line flux. A full version of this table, including an expanded set of emission line identifications and measurements, can be found online.
\end{table*}


\begin{figure}[!t]
    \centering
    \includegraphics[width=9cm]{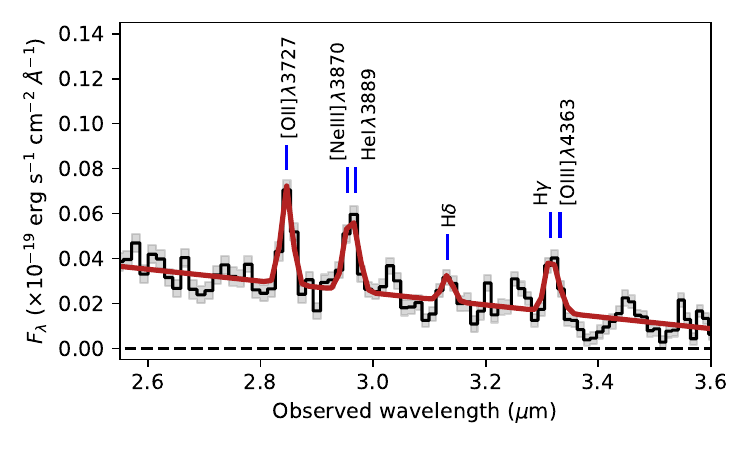}
    \includegraphics[width=9cm]{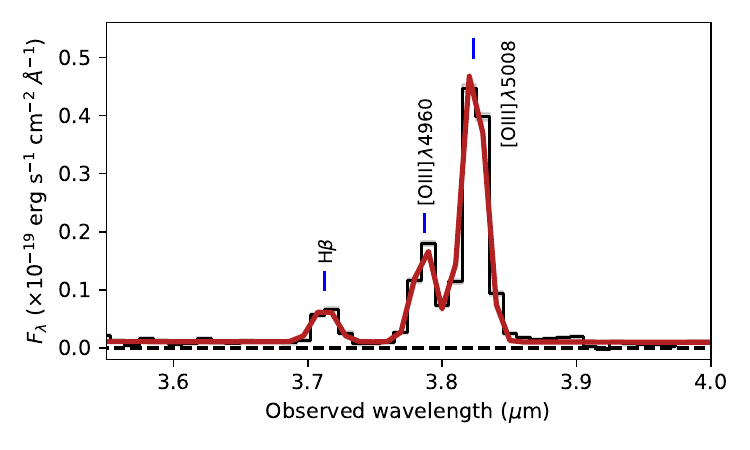}
    \includegraphics[width=9cm]{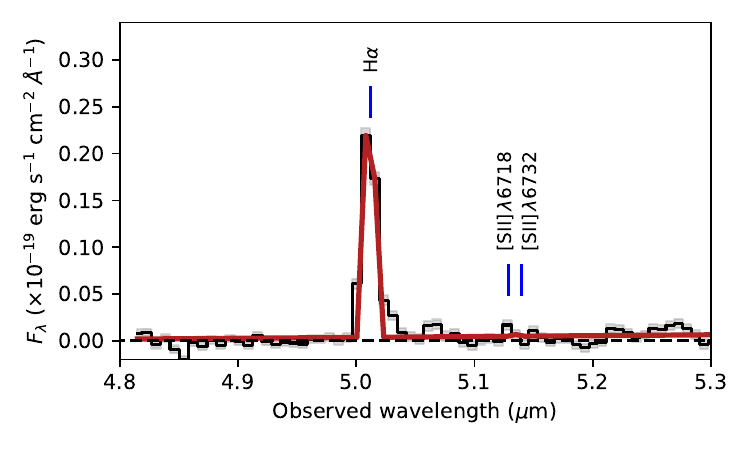}
    \caption{Example of line flux modeling for one of the PRIMAL sources at $z=6.6345$. The \jw/NIRSpec Prism spectrum is shown in black and the associated error spectrum in gray. The best fit continuum and Gaussian line model is shown in red. }
    \label{fig:lineflux}
\end{figure}

\subsection{Line flux measurements}

For each source in the \jw-PRIMAL sample, we measure the spectroscopic redshifts from the most prominent nebular emission lines, when detected, including: the [\oii]\,$\lambda\lambda 3726,3729$ and [\oiii]\,$\lambda\lambda 4960,5008$ doublets, [\neiii]\,$\lambda3870$, \hei\,$\lambda 3889$, the [\sii]\,$\lambda\lambda 6718, 6732$ doublet, and the Balmer lines H$\alpha$, H$\beta$, H$\gamma$, and $H\delta$. The [\oii]\,$\lambda\lambda 3726,3729$ and [\sii]\,$\lambda\lambda 6718, 6732$ doublets are unresolved at all wavelengths in the NIRSpec Prism spectra, whereas the [\oiii]\,$\lambda\lambda 4960,5008$ doublet is marginally resolved at $z\approx 6$ and fully resolved at $z\gtrsim 7.5$ due to the increasing spectral resolution with wavelength in this particular configuration. H$\alpha$ and [\nii]\,$\lambda 6585$ are unresolved at all redshifts. [\neiii]\,$\lambda3870$ and \hei\,$\lambda 3889$ are also generally unresolved, except for the highest redshifts at $z\gtrsim 10$.
Due to the broad wavelength coverage ($0.6-5.3\mu$m), H$\alpha$ and [\sii]\,$\lambda\lambda 6718, 6732$ can be detected up to $z\approx 7$, H$\beta$ and the [\oiii]\,$\lambda\lambda 4960,5008$ doublet up to $z\approx 9.5$, and \lya\ can be detected at all redshifts from $z\gtrsim 5.5$. The auroral [\oiii]\,$\lambda 4363$ line is also in the majority of cases blended with H$\gamma$ but becomes resolved at $z\gtrsim 9.5$, enabling robust measurements of this important transition \citep{Heintz23_DLA,Hsiao23}. Unfortunately, the [\oiii]\,$\lambda\lambda 4960,5008$ doublet is redshifted out of the NIRSpec Prism coverage at approximately the same redshifts, hindering direct $T_e$-based metallicity measurements with the \jw/NIRSpec Prism configuration, except for a narrow redshift range ($z\approx 9.3-9.5$). 

To determine the physical properties of the star-forming regions in each of the target galaxies, we measure or derive upper bounds on the line fluxes for each of these nebular emission lines when covered by the spectra. We fit the continuum around the lines with a simple polynomial and superimpose a set of redshifted Gaussian line profiles at the rest-frame wavelength of each transition and fit for the redshift $z_{\rm spec}$, line equivalent widths (EWs) and fluxes. The line widths (full-width-at-half-maximum; FWHM) are in most cases directly proportional to the line-spread function of the NIRSpec Prism configuration \citep{Jakobsen22} and is modelled as such. For each case, we tie $z_{\rm spec}$ and the FWHM corrected by the spectral resolution, effectively assuming that the nebular emission lines all originate from the same \hii\ regions. An example of the line model fits is shown in Fig.~\ref{fig:lineflux}. All the main identified lines, their derived line fluxes, or $2\sigma$ upper limits, are summarized in Table~\ref{tab:lineflux} for the \jw-PRIMAL sample and provided in full online\footnote{\url{https://github.com/keheintz/jwst-primal}}, which also includes an expanded line list. 

\subsection{[\oiii]+H$\beta$ equivalent widths}

\begin{figure}[!t]
    \centering
    \includegraphics[width=9cm]{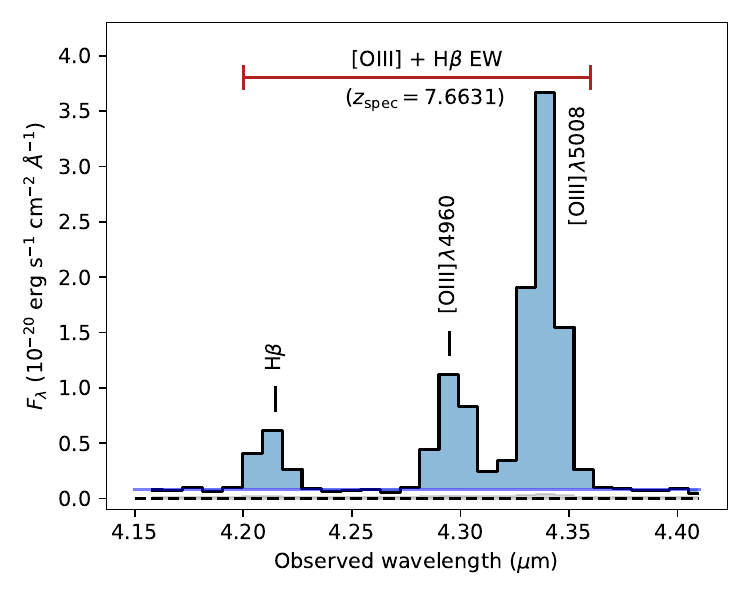}
    \caption{Example of a [\oiii]+H$\beta$ line equivalent width measurement. The blue-shaded region marks the lines integrated over, also represented by the top red bar. The continuum flux is measured by fitting a linear polynomial to the spectrum on each side of the lines.}
    \label{fig:o3hbewmeas}
\end{figure}

A simple diagnostic of the specific star-formation rate (sSFR) of galaxies is the emission line [\oiii]+H$\beta$ EWs. These have been found to increase with increasing redshifts \citep{Khostovan16,Endsley21,Matthee23}, and toward lower stellar masses and metallicities \citep{Malkan17,Reddy18a}, due to the more intense ionization fields and active star formation in early galaxies. We derive the [\oiii]+H$\beta$ EW for each of the sources in our sample that have these lines covered by the spectra (311 sources at $z<9.5$). As demonstrated in Fig.~\ref{fig:o3hbewmeas}, we integrate the flux density over the spectral region covering the H$\beta$ and [\oiii]\,$\lambda\lambda 4960,5008$ lines, normalized by the continuum flux, and correct the redshifted EWs to the rest frame. For each source, we extrapolate the continuum flux at the position of the lines by fitting a linear polynomial to the spectral regions on each side of the line complex. 

\begin{figure}[!t]
    \centering
    \includegraphics[width=9cm]{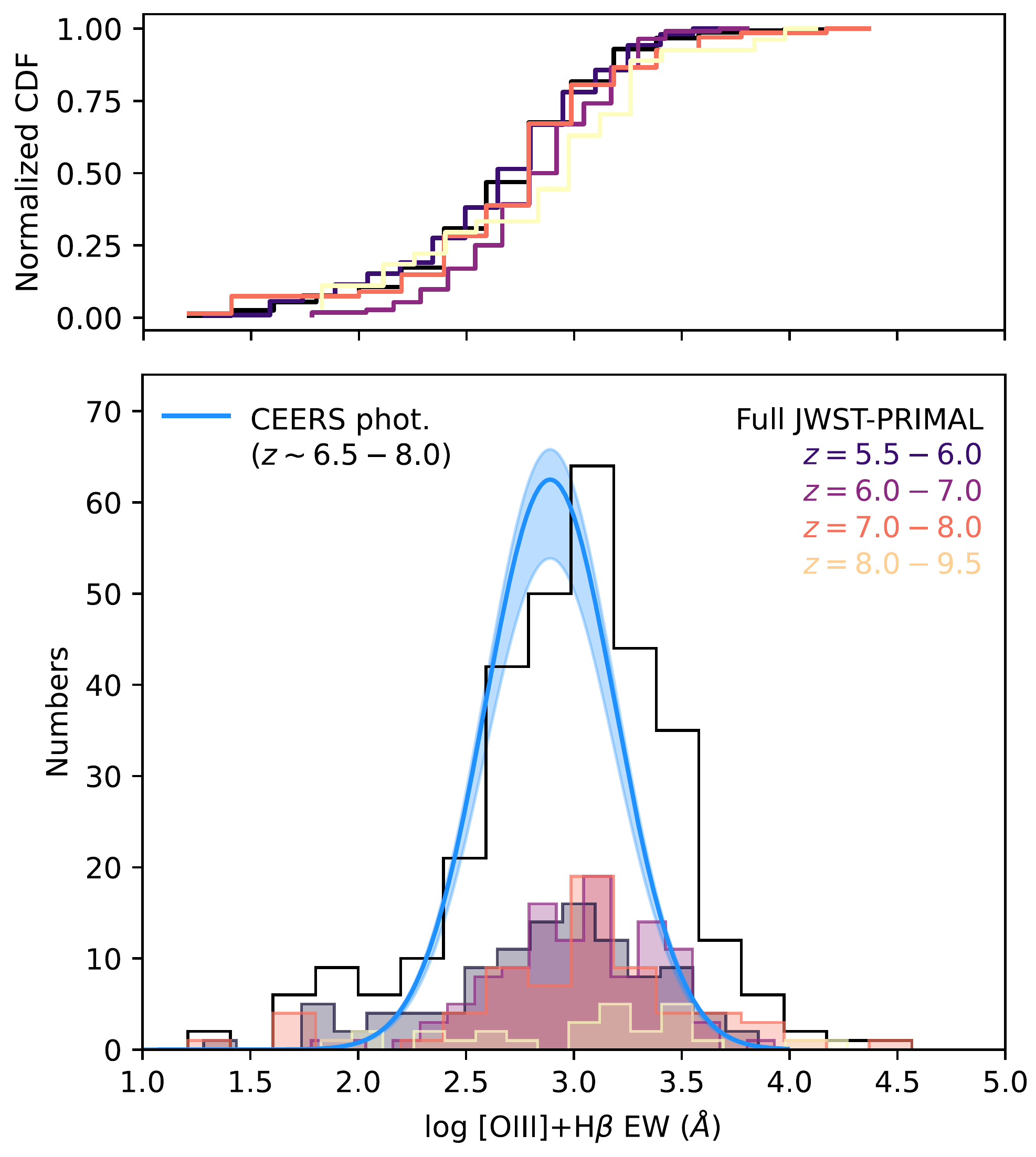}
    \caption{[\oiii]+H$\beta$ emission line equivalent width distribution (bottom) and normalized cumulative distribution function (top). The full sample up to $z\approx 9.5$, where these lines can be detected, is shown by the black step function, and divided into redshift bins as indicated by the colors. The observed distribution of the spectroscopic \jw-PRIMAL sample is compared to the larger photometric CEERS survey results from \citet{Endsley23}.}
    \label{fig:o3hbewdist}
\end{figure}

In Fig.~\ref{fig:o3hbewdist}, we show the [\oiii]+H$\beta$ EW distribution of the full \jw-PRIMAL sample and divided into redshift bins: $z=5.5$--6.0, $z=6.0$--7.0, $z=7.0$--8.0, and $z=8.0$--9.5. The median of the full distribution is $1045\,\AA$, with 16th and 84th distribution percentiles from $356\,\AA$ to $2543\,\AA$. We observe no apparent evolution with redshift, but note that the number statistics is largely dominated by the lowest redshift bins. We also compare the full distribution to the much larger, albeit photometrically derived distribution from the CEERS survey of galaxies at $z=6.5$--8.0 \citep{Endsley23}. The two distributions are in good agreement, though the peak of the spectroscopic distribution is slightly shifted towards higher EWs. We note that the spectroscopically derived [\oiii]+H$\beta$ EWs are much more accurate than photometric estimates, however, the latter potentially being underestimated by up to $\approx 30\%$ \citep{Duan23}. The fact that EW distribution of the \jw-PRIMAL sample is only slightly shifted, but otherwise follow the same distribution, indicates that there are no strong biases in our spectroscopic sample toward strong emission-line sources from the underlying galaxy population. 

\subsection{Ultraviolet spectral slopes, magnitudes and luminosities} \label{ssec:uv}

The UV spectral slope, $\beta_{\rm UV}$, where $F_\lambda \propto \lambda^{\beta_{\rm UV}}$, of the stellar continuum at wavelengths $\lambda_{\rm rest} = 1250$--2600\,$\AA$ encodes key information regarding dust attenuation \citep{Calzetti94,Meurer99}, stellar metallicity \citep{Cullen21}, the average stellar population age~\citep{Zackrisson11}, and the escape fraction of ionizing photons \citep{Chisholm22} from galaxies. The minimum value expected for standard stellar populations with a Galactic IMF and with negligible dust attenuation is $\beta_{\rm UV} \approx -2.5$ \citep{Cullen17,Reddy18b}. Steeper spectral power-law indices of $\beta_{\rm UV} \approx -3.0$ would require a very recent onset of star formation ($\lesssim 2\,$Myr) or near-pristine, neutral gas and negligible nebular processing. \jw has enabled measurements of $\beta_{\rm UV}$ of galaxies at $z>10$, hinting at even steeper spectral slopes, particularly for the faintest, less massive systems \citep{Topping23,Cullen23a}. These results have, however, been inferred from photometric data alone, which may be contaminated by varying \lya\ or metal emission (or absorption) line strengths or even low-luminosity AGN.

\begin{figure}[!t]
    \centering
    \includegraphics[width=9.2cm]{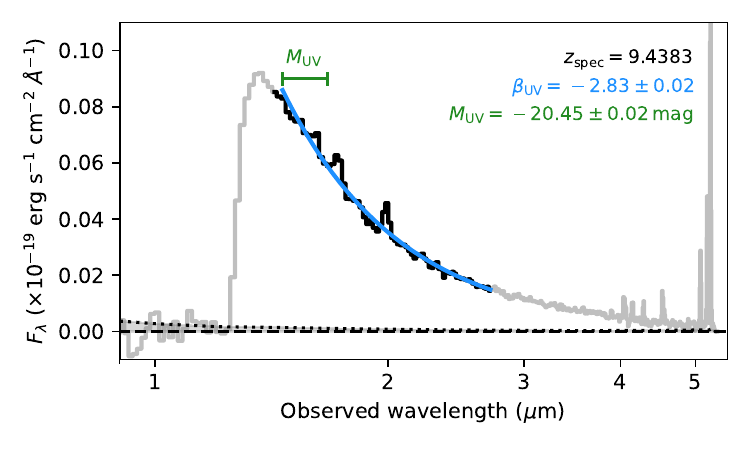}
    \caption{Example of spectral fitting of the UV power-law index, $\beta_{\rm UV}$, where $F_\lambda \propto \lambda^{\beta_{\rm UV}}$, of the stellar continuum. The grey curve shows the full NIRSpec Prism spectrum of a galaxy at $z=9.4383$, with the error spectrum shown by the dotted line. The black part of the spectrum highlights the rest-frame fitting region from $\lambda_{\rm rest} = 1250-2600\,\AA$. The top green bar marks the integration region for the $M_{\rm UV}$ measurements, $\lambda = 1400-1700\,\AA$ in the rest frame.}
    \label{fig:betafit}
\end{figure}

\begin{figure}[!ht]
    \centering
    \includegraphics[width=9.2cm]{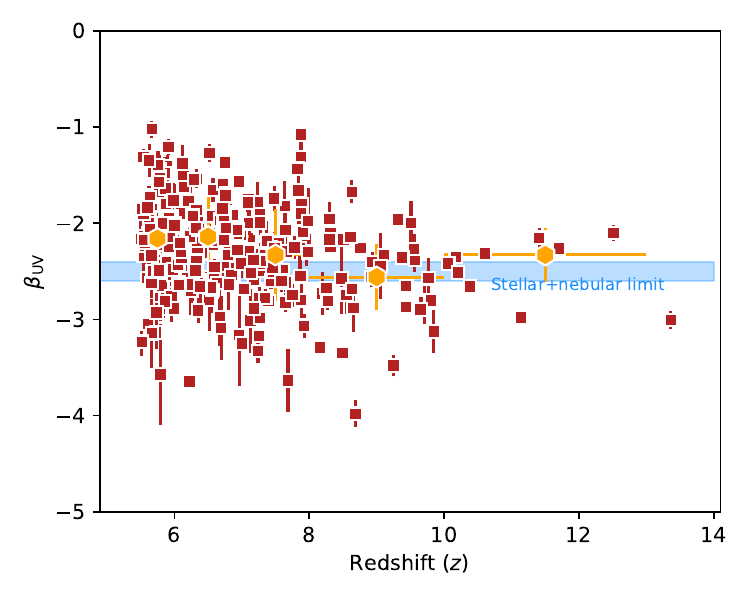}
    \caption{The galaxy UV spectral slope, $\beta_{\rm UV}$ as a function of redshift.  The red points mark the \jw-PRIMAL sample from this work, and the shown quantities are all derived directly from the NIRSpec Prism spectroscopy. The yellow hexagons show the mean $\beta_{\rm UV}$ in bins denoted by the horizontal errorbars, and the vertical errorbars represent the 16th to 84th percentile of the distribution in the respective bins. The blue shaded region represents the minimum value $\beta_{\rm UV} = -2.6$ to $-2.4$ for a standard stellar population with negligible dust and maximum nebular continuum contribution \citep[see Sect.~\ref{ssec:uv} and][]{Cullen23b}.}
    \label{fig:betaz}
\end{figure}

Here, we measure $\beta_{\rm UV}$ directly from the spectra for each of the \jw-PRIMAL sources at $z=5.5$--13.4, carefully masking any identified emission lines or broad \lya\ absorption troughs. We use the {\tt emcee} minimizer within the {\tt LMFit} framework and recover the parameter covariances to estimate the median and 16th to 84th percentiles of the posterior distributions on $\beta_{\rm UV}$ and the normalization constant derived at UV $1550\,\AA$ rest-frame, $m_{\rm UV}$ of the best-fit model, $F_\lambda = F_{\rm UV,1550} \lambda^{\beta}$. An example of one of the model fits is shown in Fig.~\ref{fig:betafit}. For our full sample, we recover UV continuum slopes ranging from $\beta_{\rm UV} =-1$ down to a few cases with $\beta_{\rm UV} \lesssim -3.0$. This suggests that this set of rare galaxies are particularly dust- and metal-poor ($Z_{\rm gas}/Z_\odot < 1\%$), have mean stellar populations with ages $\lesssim 2$\,Myr \citep[e.g.,][]{Cullen23b}, and likely have ionizing photon escape fractions of $\gtrsim 10\%$ \citep{Chisholm22}. In Fig.~\ref{fig:betaz}, we show the derived $\beta_{\rm UV}$ spectral slopes as a function of redshift. We observe a marginal tendency for galaxies to have steeper UV slopes at higher redshifts, though still consistent with a non-evolution within the scatter in the sample distributions. We further note that distribution of dust and \hii regions may account for the large scatter in the population around the expected slope for a standard stellar population with negligible dust and maximum nebular continuum contribution \citep[e.g.,][]{Calzetti94,Vijayan24}.

To determine the UV luminosity, $L_{\rm UV}$, and absolute magnitude, $M_{\rm UV}$, for each source we derive the flux density and corresponding UV magnitude, $m_{\rm UV}$ at rest-frame $1550\,\AA$ from the spectral power-law model. We convert the apparent $m_{\rm UV}$ to the intrinsic, absolute brightness, $M_{\rm UV}$, via
\begin{equation}
    M_{\rm UV} = m_{\rm UV} - \mu(z) + 2.5 \times \log (1+z)
\end{equation}
where $\mu(z)$ is the distance modulus at a given redshift derived from the {\tt astropy.cosmology} module.
The derived $M_{\rm UV}$ for each source are summarized in Table~\ref{tab:physprop}, not corrected for potential magnification factors for galaxies drawn from lensing cluster surveys. Our full sample spans absolute UV magnitudes of $M_{\rm UV} = -22$ to $-16$\,mag, respectively (see Fig.~\ref{fig:muvzsp}).

\begin{table*}
\begin{center}    
\caption{Spectroscopically derived physical properties of the \jw-PRIMAL sample.} 
\label{tab:physprop}      
\setlength\tabcolsep{0.32cm}
\begin{tabular}{c c c c c c c c c}
\hline\vspace{0.1cm}
$z_{\rm spec}$ & $M_{\rm UV}$& [\oiii]+H$\beta$ EW & $\beta_{\rm UV}$ & $D_{\rm Ly\alpha}$ & SFR$_{\rm H\beta,[OII]}$ & $O_{32}$ & $12+\log({\rm O/H})$ \\ 
& (mag) & ($\AA$) & & ($\AA$) & ($M_\odot$\,yr$^{-1}$) &  & \\
 (1) &  (2) &  (3) &  (4) & (5) & (6) & (7) & (8)  \\
\hline   
5.5050 & $-18.83$ & $1612.3$ & $-2.16\pm 0.06$ & $52.3\pm 2.3$ & $4.1\pm 0.2$ & $8.6\pm 1.0$ & $7.95\pm 0.15$ \\
5.5050 & $-18.85$ & $2850.3$ & $-1.69\pm 0.26$ & $63.2\pm 14.8$ & $1.16\pm 0.12$ & $10.71\pm 4.60$ & $7.35\pm 0.15$ \\
5.5161 & $-19.04$ & $1383.2$ & $-3.23\pm 0.15$ & $-90.7\pm 6.5$ & $1.77\pm 0.12$ & $15.00\pm 6.43$ & $7.28\pm 0.15$ \\
5.5193 & $-16.83$ & $20235.6$ & $-2.30\pm 0.32$ & $42.5\pm 6.5$ & $ 3.12\pm 0.92$ & $9.41\pm 4.12$ & $7.90\pm 0.15$ \\
5.5207 & $-20.03$ & $671.72$ & $-1.85\pm 0.02$ & $40.6\pm 4.2$ & $1.78\pm 0.18$ & $2.52\pm 0.16$ & $7.76\pm 0.15$ \\
$\vdots$ & $\vdots$ & $\vdots$ & $\vdots$ & $\vdots$ & $\vdots$ & $\vdots$ & $\vdots$ \\
11.4906 & $-19.96$ & $-$ & $-1.79\pm 0.13$ & $35.6\pm 7.1$ &  $-$ &  $-$ &  $-$  \\
11.7060 & $-19.57$ & $-$ & $-2.26\pm 0.06$ & $48.4\pm 2.6$ &  $2.90\pm 0.72$ &  $-$ &  $8.34\pm 0.24$ \\
12.5119 & $-18.96$ & $-$ & $-2.10\pm 0.08$ & $90.4\pm 3.9$ & $0.37\pm 0.18$ &  $-$ &  $7.91\pm 0.24$ \\
12.7822 & $-19.57$ & $-$ & $-2.53\pm 0.01$ & $32.1\pm 7.3$ &  $<2.15$ &  $-$ &   $-$ \\
13.3605 & $-19.40$ & $-$ & $-3.01\pm 0.09$ & $56.7\pm 3.1$ & $1.47\pm 0.49$ &  $-$ &   $-$ \\
\hline          
\end{tabular} 
\end{center}  
\textbf{Notes.} Column (1): Spectroscopic redshift. Column (2): Absolute UV (rest-frame $\approx 1500\,\AA$) magnitude. Column (3): Rest-frame [\oiii]+H$\beta$ equivalent width. Column (4): Rest-frame UV ($1250-2600\,\AA$) spectral slope. Column (5): The \lya\ damping parameter. Column (6): The star-formation rate derived from H$\beta$ ($z<10$) or [\oii] ($z>10$). Column (7): The [\oiii]\,$\lambda 5008 / $ [\oii]\,$\lambda 3727$ line flux ratio. Column (8): Gas-phase metallicity inferred from the joint PDF of the different strong-line calibrations from \citet{Sanders24}. A full version of this table can be found online.
\end{table*}

\subsection{Star-formation rate}

The Balmer recombination lines originating from star-forming \hii\ regions, H$\alpha$ in particular, are robust tracers of star formation on short ($\lesssim 20$\,Myr) timescales. The star-formation rate (SFR) has been found to increase at a given stellar mass \citep{Whitaker12,Speagle14,Salmon15,Thorne21,Sandles22} and rest-frame UV size \citep{VanDerWel14,Ward23,Langeroodi23b} for galaxies at increasing redshifts, following the evolution in the peak of the stellar mass halo mass relation with redshift \citep[e.g.,][]{Behroozi19}. The SFR can now be measured directly from the Balmer recombination lines for galaxies at $z>6$ with \jw, revealing intense star-forming galaxies during the reionization epoch \citep[e.g.,][]{Heintz23_FMR,Shapley23,Fujimoto23_ceers,Sanders24,Nakajima23,Curti23b}.
Given that H$\alpha$ is only detected for the subset of the sample at $z<7$, and may be contaminated by [\nii], we instead derive the SFR based on H$\beta$ for the majority of sources at $z<10$ as: 
\begin{equation}
    {\rm SFR_{H\beta}} (M_\odot\,{\rm yr}^{-1}) = 5.5\times 10^{-42} L_{\rm H\beta} ({\rm erg\,s^{-1}}) \times f_{\rm H\alpha / H\beta} ~~ (z<10)
\end{equation}
following \citet{Kennicutt98}, but here assuming the more top-heavy initial mass function (IMF) from \citet{Kroupa01}, which is likely more representative of the high-redshift galaxy population \citep[e.g.,][]{Steinhardt23}. We derive the H$\beta$ line luminosity via $L_{\rm H\beta} = F_{\rm H\beta} \times 4\pi D^2_L(z)$ with $D_L(z)$ the luminosity distance at the given redshift and assume $f_{\rm H\alpha / H\beta} = 2.86$ from the Case B recombination scenario at $T_e = 10^{4}$\,K \citep{Osterbrock06}. We propagate the uncertainty of $\approx 20$--30\% from the exact choice of the IMF in the relation, in addition to the dependence on the electron density and temperature of the \hii\ region which is only of the order $\lesssim 5\%$. 

For the sources at $z>10$, where H$\beta$ is redshifted out of the NIRSpec Prism spectral coverage, we instead rely on the [\oii]\,$\lambda 3727$ line strength
\begin{equation}
    {\rm SFR_{[OII]}} (M_\odot\,{\rm yr}^{-1}) = 1.0\times 10^{-41} L_{\rm [OII]} ({\rm erg\,s^{-1}})  ~~ (z>10)
\end{equation}
again following \citet{Kennicutt98}, and assuming the \citet{Kroupa01} IMF. For the full sample, we derive SFRs in the range 0.1--$100\,M_\odot\,{\rm yr}^{-1}$, not corrected for the relevant magnification factors or dust extinction, as summarized in Table~\ref{tab:physprop}. The SFRs can also be derived from overall UV luminosity of the sources following \citet{Madau14}, where 
\begin{equation}
    {\rm SFR_{UV}} (M_\odot\,{\rm yr}^{-1}) = 1.0\times 10^{-28} L_{\rm UV} ~ ({\rm erg\,s^{-1}\,Hz^{-1}})
\end{equation}
assuming $10\%$ solar metallicity which traces star formation on a longer timescale ($\sim 100$\,Myr) compared to the Balmer line \citep[$\sim 100$\,Myr, see][]{Calzetti13}. We find that the ${\rm SFR_{UV}}$ estimate is consistent with that derived from the Balmer recombination lines or [\oii] for the full sample, when allowing for up to $A_V\approx 0.5$\,mag of dust attenuation. 



\subsection{Gas-phase metallicity}

One of the most fundamental characteristics of galaxies is their metallicity. This is notoriously difficult to measure directly, in particular for the highest redshift galaxies, motivating the use of strong-line diagnostics of the most prominent nebular emission lines \citep[see e.g.][for recent reviews]{Kewley19,Maiolino19}. In rare cases, the auroral [\oiii]\,$\lambda 4363$ emission line can be detected directly, which enables $T_e$ sensitive measurements of the gas-phase metallicity (known as the direct $T_e$-method), typically quantified as the oxygen abundance, $12+\log{\rm (O/H)}$ \citep[e.g.,][]{Izotov06}. This particular feature has now been detected in star-forming galaxies at $z>6$ with \jw \citep[e.g.,][]{Schaerer22,Curti23a,Heintz23_JWSTALMA,Nakajima23,Sanders24}, even out to $z\approx 10$ \citep{Hsiao23,Heintz23_FMR,Heintz23_DLA}. The auroral [\oiii]\,$\lambda 4363$ line feature, however, will only be resolved from H$\gamma$ in the minority of sources due to the increasing spectral resolution towards the red end of the spectra. 

\begin{figure}[!t]
    \centering
    \includegraphics[width=9cm]{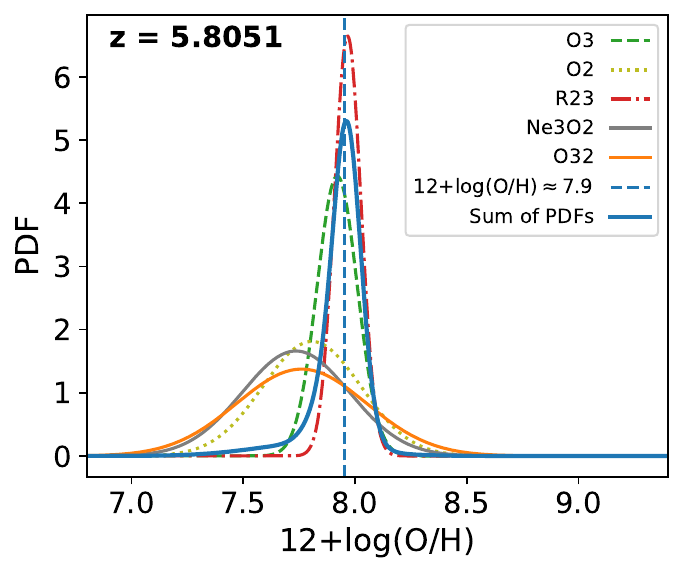}
    \caption{Example of joint gas-phase metallicity estimate for one of the sources. The blue-colored function shows the normalized joint probability density function (PDF) based on the normalized PDFs of the individual line-ratio diagnostics from \citet{Sanders24}. The blue dashed line marks the median of the joint PDF.}
    \label{fig:logoh}
\end{figure}


We therefore base our main metallicity measurements on strong-line calibrations. To preserve the homogeneity of the measurements, we use a single set of line diagnostics from \citet{Sanders24} for the different line ratios, derived for galaxies at $z=2$--9 through the direct $T_e$-method. Specifically, we consider the strong-line ratios: ${\rm O3} = $~[\oiii]\,$\lambda 5008 / {\rm H}\beta$, ${\rm O2} = $~[\oii]\,$\lambda 3727 / {\rm H}\beta$, ${\rm R23} = $~([\oiii]\,$\lambda\lambda 4960,5008$ + [\oii]\,$\lambda 3727$) / H$\beta$, ${\rm Ne3O2} = $~[\neiii]\,$\lambda 3870$ / [\oii]\,$\lambda 3727$, and ${\rm O32} = $~[\oiii]\,$\lambda 5008 / $ [\oii]\,$\lambda 3727$, which are available at most of the targeted redshifts. Based on the line-flux measurements and errors (representing the likelihood distributions of fluxes), we construct the probability density function (PDF) for each diagnostic ratio where the width denotes the scatter in the relation. We then combine the PDFs from the available calibrations, inversely weighted by their scatter, for each individual source to determine the gas-phase metallicity through the median and 16th to 84th percentiles of the joint PDF, as illustrated in Fig.~\ref{fig:logoh}. This more conservative approach takes into account the added uncertainty from sources with potential high ionization parameters, typically quantified via $O_{32}$, thus minimizing the uncertainty from assuming a single line-ratio. 

\begin{figure*}[!ht]
    \centering
    \includegraphics[width=17cm]{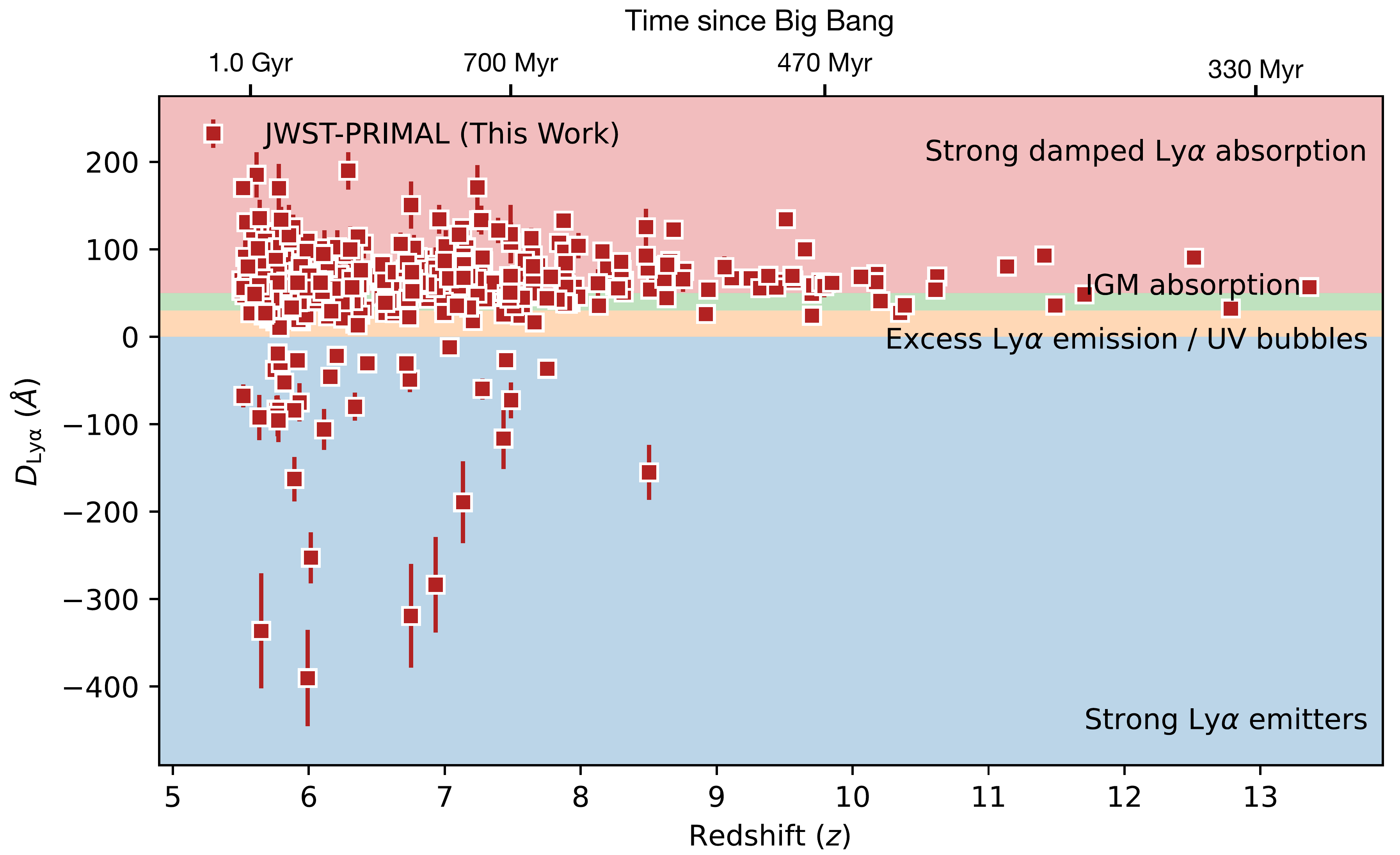}
    \caption{\lya\ damping parameter, \dlya, for the full \jw-PRIMAL sample as a function of redshift. The corresponding age of the Universe assuming the cosmological parameters from \citet{Planck18} is given at the top. The colored backgrounds indicate the typical physical regimes probed for a given \dlya: Strong \lya emission ($D_{\rm Ly\alpha} < 0\,\AA$; blue), extended ionized bubbles or weak \lya emission ($D_{\rm Ly\alpha} = 0$--$35\,\AA$; orange), IGM absorption from $x_{\rm HI} = 0$--1 ($D_{\rm Ly\alpha} = 35$--$50\,\AA$; green), and strong galaxy integrated DLAs with $N_{\rm HI} > 10^{21}\,$cm$^{-2}$ from local \hi\ gas ($D_{\rm Ly\alpha} > 50\,\AA$; red). }
    \label{fig:dlyaz}
\end{figure*}

We derive gas-phase metallicities ranging from $12+\log{\rm (O/H)} = 6.5$ to $8.2$, corresponding to 0.6\% to 30\% of solar abundance, respectively \citep[given $12+\log{\rm (O/H)_\odot} = 8.69$;][]{Asplund09}. The relevant line fluxes used for these measurements are provided in Table~\ref{tab:lineflux} and the derived metallicities are listed for each source in Table~\ref{tab:physprop}. 

\section{Lyman-$\alpha$ emission and absorption in galaxies during the reionization epoch} 
\label{sec:dlas}

With the spectroscopically-derived physical properties for the full set of \jw-PRIMAL sample sources we can now investigate and chart the redshift evolution and the physical driver of strong galaxy DLAs produced by massive \hi\ gas reservoirs to prominent ionized bubbles and \lya\ emission during the reionization epoch.      

\subsection{Charting \dlya\ as a function of redshift} 

To obtain a first overview of the prevalence and prominence of DLAs and LAEs across time in the reionization epoch, we show \dlya\ as a function of redshift for the full sample in Fig.~\ref{fig:dlyaz}. This represents the evolution from $z=13.4$ to $z=5.5$, corresponding to approx. 300~Myr to 1~Gyr after the Big Bang. The \dlya\ parameter space has been color-coded to highlight the regimes driven by the physical models for the expected dominant contribution to the observed \lya\ transmission curve. The number statistics of sources located in each region are summarized in Table~\ref{tab:stats}, divided into redshift bins of $z=5.5$--6.0 (159 sources), $z=6.0$--7.0 (169), $z=7.0$--8.0 (103), $z=8.0$--10.0 (46), and $z=10.0$--13.4 (17). 

We find that there are 38 strong LAEs in the full \jw-PRIMAL sample (i.e. $8\%$), based on $D_{\rm Ly\alpha} < 0\,\AA$. None of these are found at $z>9$. At $z<8$, we find that 38 out of 494 of sources show strong LAEs ($8\%$) in approximately equal fractions from $z=5.5$--6.0, $z=6.0$--7.0, and $z=7.0$--8.0, albeit with the most prominent LAEs quantified by the lowest \dlya\ at $z<7$. This fraction of LAEs could be as high as $\approx 20\%$ if we include all sources with excess \lya\ flux as potential weakly-emitting LAEs. These results are slightly lower than found previous literature studies, showing a typical fraction of LAEs of $\approx 30\%$ at $z\approx 6$--8 \citep[e.g.,][]{Saxena23,Nakane23,Simmonds23,Witstok23,Jung23,Jones23,Witten24}. However, they were also found to be much more sparsely populated beyond $z\gtrsim 8$ \citep[though see][Witstok et al. in prep.]{Tang23,Bunker23_gnz11,Fujimoto23_uncover}, consistent with our results, and as also recovered by simulations \citep{Hutter23}. These findings would also naturally explain the null detections of LAEs at $z=8.8$ in large photometric surveys using narrow-band imaging such as the UltraVISTA \citep[e.g.,][]{Laursen19}. 

\begin{table*}[!t]
    \begin{center}
    \caption{\lya\ absorption vs. emission statistics.}
    \label{tab:stats}
    \begin{tabular}{l ccccc}
    \hline
    {\bf Redshift range} & $5.5 - 6.0$ & $6.0-7.0$ &  $7.0-8.0$ & $8.0 - 10.0$ & $10.0 - 13.4$ \\
    {\bf LAEs} & 18/159 ($11\%$) & 12/169 ($7\%$) & 7/103 ($7\%$) & 1/46 ($2\%$) & 0/17 ($0\%$) \\
    {\bf Ly$\alpha$ exc.} & 13/159 ($9\%$) & 15/169 ($9\%$) & 4/103 ($4\%$) & 2/46 ($4\%$) & 1/17 ($6\%$)  \\ 
    {\bf IGM abs.} & 34/159 ($21\%$) & 28/169 ($17\%$) & 23/103 ($22\%$) & 3/46 ($7\%$) & 5/17 ($29\%$)  \\
    {\bf DLAs} & 94/159 ($59\%$) & 114/169 ($67\%$) & 69/103 ($67\%$) & 40/46 ($87\%$) & 11/17 ($65\%$)  \\
    \hline    
    \end{tabular}
    \end{center}
\end{table*}

\begin{figure}[!t]
    \centering
    \includegraphics[width=9cm]{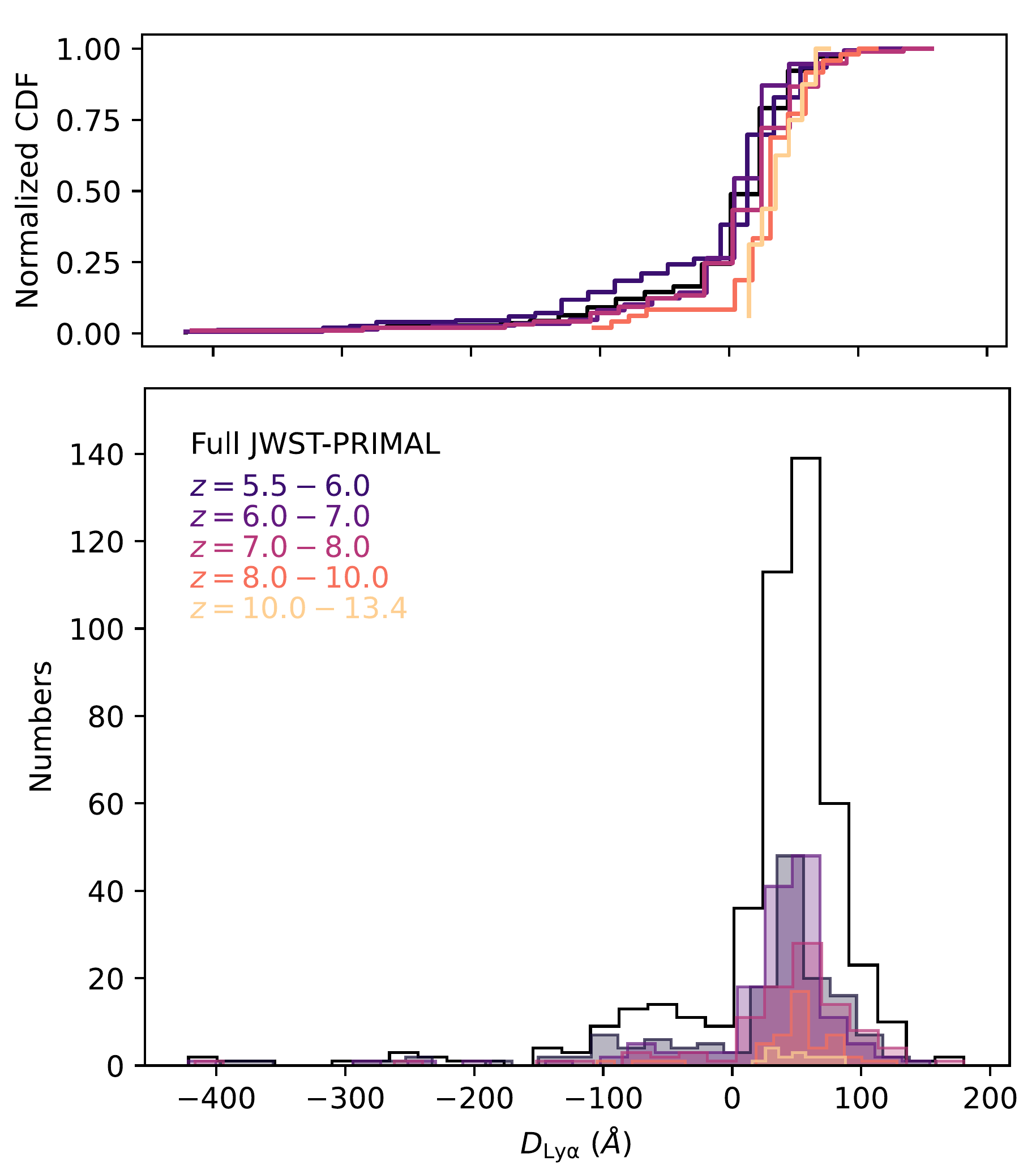}
    \caption{Histogram of the \lya\ damping parameter, \dlya, distribution (bottom) and normalized cumulative distribution function (top). The full \jw-PRIMAL sample is represented by the black step function, and divided into sub-bins according to their redshift as indicated by the colors. Strong DLAs are more prevalent in galaxy spectra at increasing redshift.}
    \label{fig:dlya_dist}
\end{figure}

Similarly, we consider the redshift evolution of strong DLAs, here classified as sources with $D_{\rm Ly\alpha} > 50\,\AA$, corresponding to \hi\ column densities $N_{\rm HI} \gtrsim 10^{21}\,$cm$^{-2}$. We find that the fraction of galaxy DLAs only slightly increases from $\approx 60\%$ at $z\approx 6$ up to $\approx 65$--90\% at $z>8$. 
A similar conclusion is reached by \citet{Umeda23}, who find an ubiquitous presence of extremely strong DLAs ($N_{\rm HI} \gtrsim 10^{22}\,$cm$^{-2}$) at $z>10$. This evolution is more clearly highlighted in Fig.~\ref{fig:dlya_dist} where we show the histograms of \dlya\ for the five different redshift ranges, demonstrating higher prevalence of galaxy-integrated DLAs at increasing redshifts. 
These results are in sharp contrast to previous studies of Lyman-break galaxies near the peak of cosmic star formation at $z\sim 2$--3 \citep{Pettini00,Steidel10}, where only a small subset of the population ($< 25\%$) show integrated galaxy line-of-sight \hi\ column densities of $N_{\rm HI} \approx 3\times 10^{20}\,$cm$^{-2}$ \citep{Shapley03}. This corroborates our findings here of an increasing probability of having high \hi\ column densities or larger fraction of DLAs at higher redshifts, down to $z\approx 2$--3. This may indicate that we are probing the epoch of first-galaxy assembly with excessive amounts of accreting or overdense pristine \hi\ gas that are yet to be processed into stars or ionized by the first stellar sources. Indeed, the average gas mass density of galaxies have been inferred to increase by an order of magnitude through accretion from $z=10$ to $6$ \citep[see][and Sect.~\ref{sec:disc} for further discussion]{Walter20,Heintz22}. (see Sect.~\ref{sec:disc} for further discussion). Fig.~\ref{fig:dlya_dist} also demonstrates an excess of galaxies with strong LAEs ($D_{\rm Ly\alpha} < 00\,\AA$) at $z=5.5$--6.0 from the full distribution, distinct for the epoch when the universe is fully ionized. 

\begin{figure}[!t]
    \centering
    \includegraphics[width=9cm]{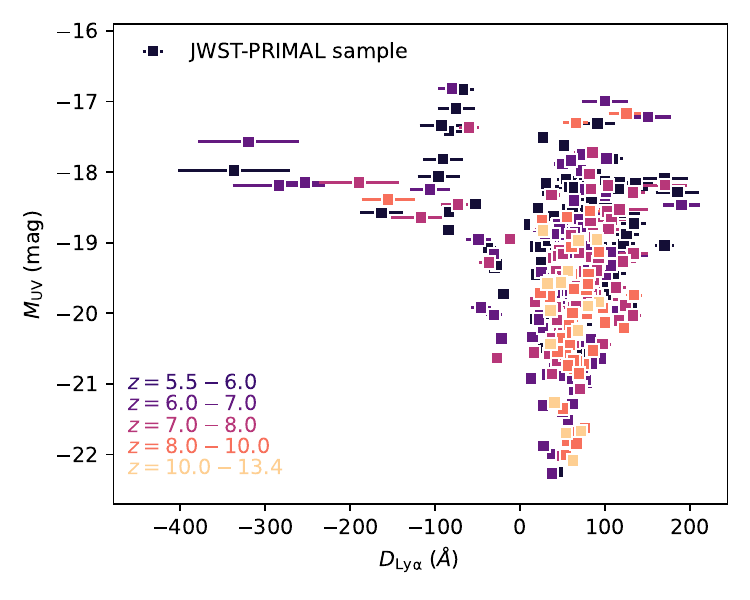}
    \caption{The absolute UV brightness, $M_{\rm UV}$, of the \jw-PRIMAL sources as a function of \dlya. The sources with the highest \lya\ EWs (negative \dlya) are observed to represent the UV faintest galaxy population at $z>5.5$.}
    \label{fig:dlya_muv}
\end{figure}

We note that the fraction of $\gtrsim 70\%$ galaxy DLAs at $z>9$ may be a lower bound, given that our selection criteria on the S/N$_{\rm Ly\alpha}$ will preferentially be biased towards strong \lya\ emission. On the other hand, fainter, less massive sources are likely to have stronger \lya\ EWs \citep{Saxena23b} and contribute to most of the ionizing photons \citep{Lin24}. To investigate this, we compare \dlya\ to the spectroscopically-derived $M_{\rm UV}$ in Fig.~\ref{fig:dlya_muv}. We confirm the trend of the intrinsically faintest sources showing the highest \lya\ EWs, with the strongest LAEs with $D_{\rm Ly\alpha} < -100\,\AA$ showing luminosities predominantly at $M_{\rm UV} > -19$\,mag. This further suggests that the large \hi\ gas overdensities are preferentially located close to bright, massive sources. 
In the following sections, we will further investigate the underlying physical properties driving the prevalence and prominence of strong \lya\ emission and absorption in these reionization era sources.



\begin{figure*}[!th]
    \centering
    \includegraphics[width=17cm]{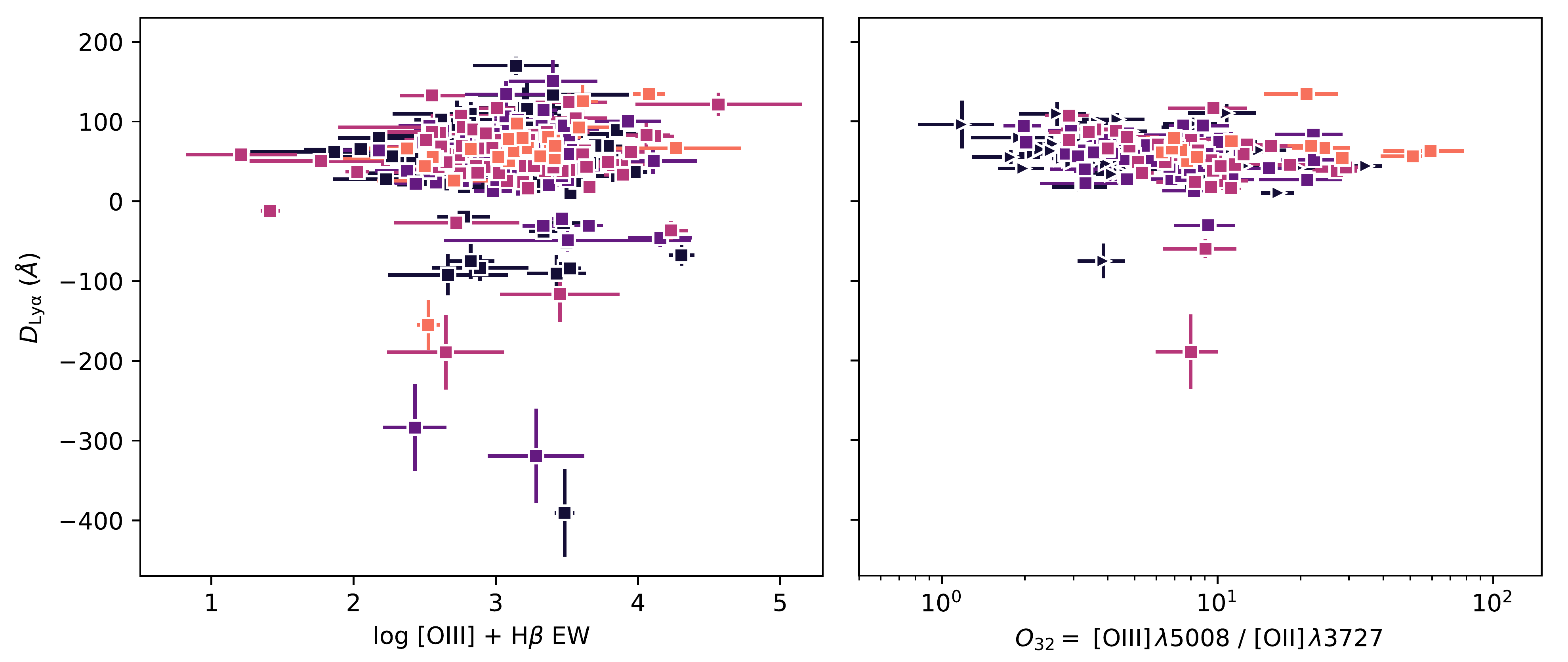}
    \caption{Correlations of \dlya\ with the galaxy properties that represent the average the specific SFR ([\oiii]+H$\beta$ EW, left), and the intensity of the stellar ionization field ($O_{32}$ = [\oiii]\,$\lambda 5008$ / [\oii]\,$\lambda 3727$, right). The square symbols show the \jw-PRIMAL sources at $z<9.5$, where [\oiii] can be detected, and a color-coded according to their redshift (identical to Fig.~\ref{fig:dlya_muv}).}
    \label{fig:dlya_emi}
\end{figure*}

\subsection{The role of stellar ionization fields on \lya\ emission and absorption}

The intense UV radiation fields produced by younger, more massive stars are likely to play a key role in the ionization of \hi\ and the production and escape of \lya\ photons. To test this and quantify it in terms of our observations, we compare \dlya to two common tracers of active star formation and high ionization: The [\oiii]+H$\beta$ EW, and the ionization parameter $O_{32}$ = [\oiii]\,$\lambda 5008$ / [\oii]\,$\lambda 3727$. We find no apparent correlations between \dlya and any of the two quantities. For example, strong galaxy DLAs are observed in galaxies spanning the entire dynamical range of  [\oiii]+H$\beta$ EWs and $O_{32}$ ionization. Similarly, strong LAEs are observed in galaxies spanning [\oiii]+H$\beta$ EWs of $100-10^{4}\,\AA$, and are not directly associated with sources with high or low $O_{32}$ ratios. This is at odds with studies of local ``green pea'' galaxies at $z\approx 0$, believed to be robust analogs to high-$z$ star-forming galaxies, that find that larger \lya\ emission EWs and escape fractions are commensurate with higher, more intense stellar ionization fields \citep{Yang17,Jaskot19,Hayes23b}. We note, however, that the galaxies at $z>8$ predominantly show high \dlya absorption and $O_{32}$ ionization and instead typically represent the large-scale environment or \hi\ gas covering fraction at these redshifts. 
Consequently, our results disfavor a scenario where the \hi\ abundance (or the lack thereof) is driven by the ionization output of the stellar sources in the galaxy.

\subsection{Gas-phase metallicity and \lya\ absorption and emission}

The gas-phase metallicity of galaxies is one of the key regulators of the gas cooling efficiency, the ionization state of the gas, and the escape fraction of \lya and LyC photons. Lower metallicities are often associated with higher ionization and temperatures, effectively reducing the observed line-of-sight \hi\ column density. By contrast, high metallicity and likely thereby also high dust content, will enable less efficient ionization of \hi\ and as a consequence more efficiently scatter the output ionizing photons, effectively suppressing the intrinsic and observed escape fraction of \lya\ photons \citep{Dayal11,Henry15,Dijkstra16,Verhamme17,Yang17,Izotov18,Jaskot19}. We  therefore expect a strong correlation between the gas-phase metallicity and \dlya\ for our high-redshift sample sources.

\begin{figure}[!t]
    \centering
    \includegraphics[width=9cm]{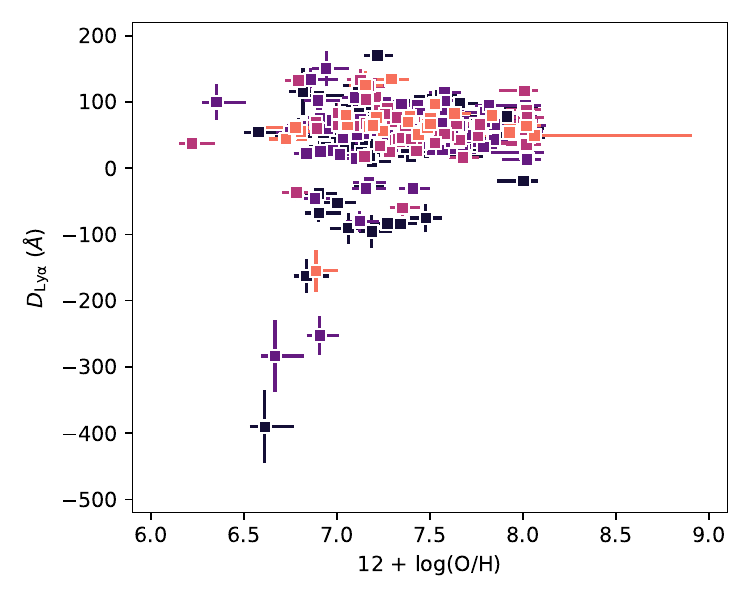}
    \caption{The gas-phase metallicity, $12+\log$(O/H), as a function of the \lya\ damping parameter, \dlya. The strength of the \lya\ absorption is observed to increase with increasing metallicity, whereas \lya\ emission is most efficiently produced or allowed to escape in low-metallicity galaxies.  }
    \label{fig:dlya_logoh}
\end{figure}

To test this scenario, we show \dlya\ as a function of gas-phase metallicity, $12+\log$(O/H), for each of the sources in the \jw-PRIMAL sample in Fig.~\ref{fig:dlya_logoh}. The strongest LAEs ($D_{\rm Ly\alpha} < 0\,\AA$) are predominantly observed in galaxies with low metallicities, and ubiquitously found at $12+\log{\rm (O/H)}<7.7$, corresponding to $<10\%$ solar metallicity. These results are consistent with the expectations that prominent \lya\ emission can only escape from low-metallicity systems. By contrast, we find that galaxies with strong DLAs ($D_{\rm Ly\alpha} > 50\,\AA$) span the entire dynamical range of inferred metallicities, $12+\log{\rm (O/H)}=6.5$--$8.2$ (i.e. 0.6--30\% solar). The most prominent DLA systems are found at lower metallicities at any redshifts $z\gtrsim 7$, supporting the scenario where these mostly trace young, less chemically enriched galaxies. 


\begin{figure*}[!t]
    \centering
    \includegraphics[width=17cm]{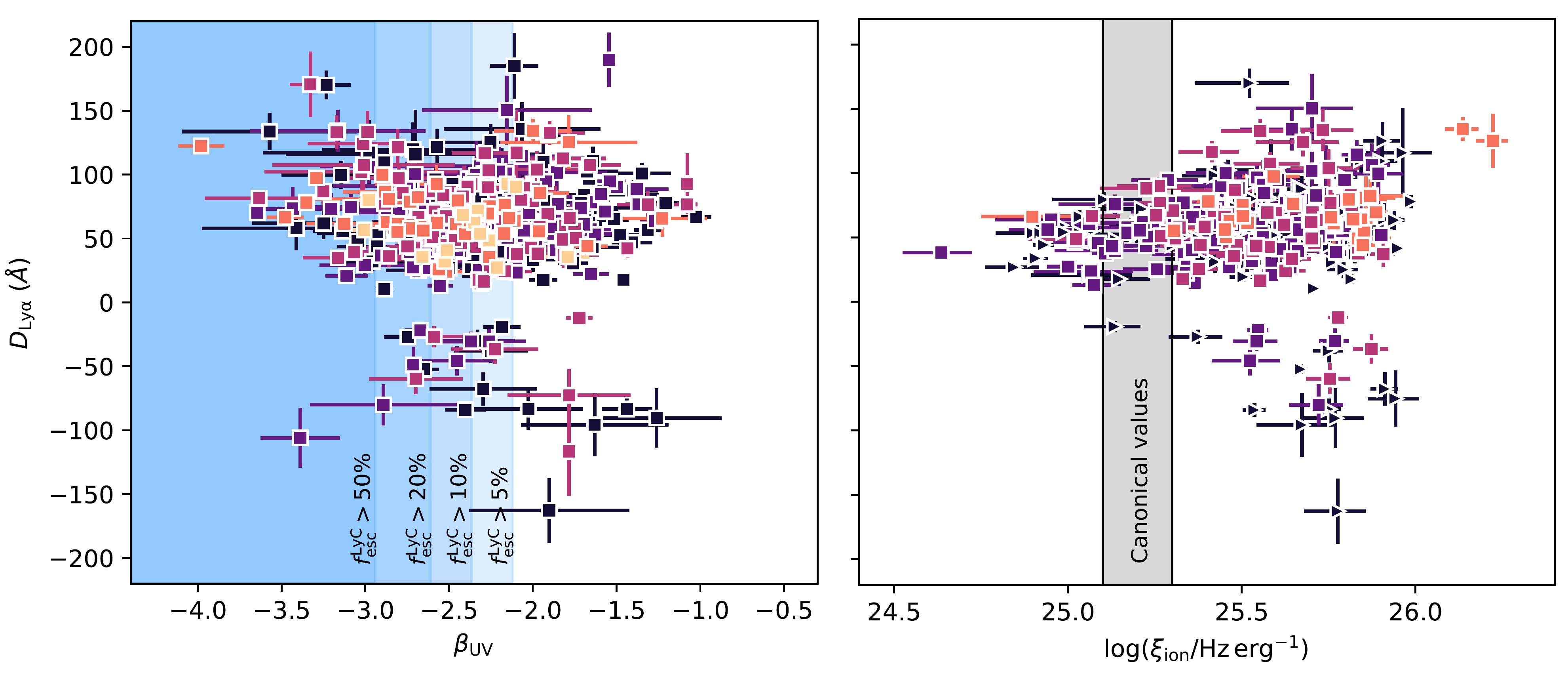}
    \caption{Comparison of \dlya\ to the spectral $\beta_{\rm UV}$ slope (left) and the ionizing photon production efficiency, $\xi_{\rm ion}$ (right). The \jw-PRIMAL sources are shown by the red squares and color-coded according to redshift (identical to Fig.~\ref{fig:dlya_muv}). The light- to dark-blue colored regions in the left panel indicates increasing escape fractions of LyC photons, $f^{\rm Lyc}_{\rm esc}$, based on the low-redshift empirical relation between $\beta_{\rm UV}$ and $f^{\rm Lyc}_{\rm esc}$ from \citet{Chisholm22}. The grey-colored band in the right panel shows the canonical reionization values $\log (\xi_{\rm ion} / {\rm Hz\,erg^{-1}}) = 25.1-25.3$ \citep{Robertson13}. Strong LAEs are observed with a large variety of $\beta_{\rm UV}$ slopes, but are predominantly associated with galaxies with high ionizing photon production efficiencies. Galaxies with strong DLAs span the entire probed range in $\xi_{\rm ion}$.}
    \label{fig:dlya_xiion}
\end{figure*}

\subsection{The impact of galaxy DLAs on the reionization history} \label{ssec:dlyaion}

One of the major implications of high \hi-cloumn density DLAs observed in galaxy-integrated spectra at $z\gtrsim 6$ is their effect on the escape fraction of \lya\ and LyC photons, key to constrain the reionization history and topology. 
Numerical modelling and reionization simulations have previously only taken IGM absorption into account when predicting the radiative transfer and transmission of \lya \citep[e.g.,][]{Laursen19,Hutter23}. These overabundant local \hi\ gas reservoirs are likely to be the main component in the transfer of ionizing photons, at least in the early evolutionary stages of galaxies, so quantifying this additional local ISM or CGM absorption is vital to draw robust conclusions and physically interpret the output and impact of galaxy-wide DLAs in simulations. 

To gauge this effect, we first compare \dlya\ to the $\beta_{\rm UV}$ spectral slopes of each galaxy in Fig.~\ref{fig:dlya_xiion}. $\beta_{\rm UV}$ has been found to be strongly correlated with the escape fraction of Lyman Continuum (LyC) photons, $f^{\rm LyC}_{\rm esc}$, in low-redshift galaxies sample \citep{Chisholm22,Flury22}. We color-grade the $\beta_{\rm UV}$ parameter space with the empirical relation presented in their paper \citet[][their Eq. 11]{Chisholm22}, highlighting the inferred escape fractions from $f^{\rm LyC}_{\rm esc} = 5\%$ to $50\%$. We find no apparent correlations between $\beta_{\rm UV}$ or the inferred $f^{\rm LyC}_{\rm esc}$ with the strength of the \lya\ emission or DLA absorption. This indicates that the hardness of the UV spectral shape is not the main driver of the ionization of \hi, and that larger \hi\ abundances do not translate directly into lower escape fractions of ionizing photons. This is somewhat at odds with theoretical predictions and local observations, since both \lya and LyC photons are affected by resonant scattered from abundant interstellar or circumgalactic \hi\ gas columns \citep[e.g.,][]{Flury22,Xu22,Begley24}. We caution, however, that the relations for $f^{\rm LyC}_{\rm esc}$ derived from the local galaxy sample at $z\approx 0.3$ studied by \citet{Chisholm22} may not be universally representative for galaxies at $z\gtrsim 6$ \citep{Choustikov24,Pahl24}. Since the high-$z$ galaxy population show even more abundant \hi\ gas reservoirs than galaxies at equivalent metallicities or $O_{32}$ ionization parameters \citep{Heintz23_DLA}, the escape fraction of \lya and LyC photons is likely correspondingly lower. Indeed, comparing the sources from \citet{Chisholm22} that are mostly absorbed by the neutral \hi\ gas, the inferred LyC escape fraction could be a $\approx 10\times$ lower. 

To more directly relate the amount of photons capable of ionizing \hi\ to the presence and strength of \lya emission or DLAs, we also compare the ionizing photon production efficiency, $\xi_{\rm ion}$, to \dlya in Fig.~\ref{fig:dlya_xiion}. $\xi_{\rm ion}$ can be derived through the Balmer recombination line strengths, such as H$\beta$, since these accurately trace the ionizing photon production rate from massive stars on short ($\lesssim 10$\,Myr) timescales \citep{Bouwens16}. Following \citet{Matthee23}, we define
\begin{equation}
    \xi_{\rm ion}\, ({\rm Hz\, erg^{-1}}) = \frac{L_{\rm H\beta}\,({\rm erg\, s^{-1}})}{c_{\rm H\beta}({\rm erg}) ~ L_{\rm UV}({\rm erg\, s^{-1}\, Hz^{-1}})} 
\end{equation}
where $c_{\rm H\beta} = 4.86\times 10^{-13}$\,erg is the H$\beta$ line-emission coefficient, assuming a Case B recombination scenario with $T_e = 10^{4}$\,K \citep[e.g.,][]{Schaerer03}. We find from Fig.~\ref{fig:dlya_xiion} that galaxies with strong DLAs ($D_{\rm Ly\alpha} > 50\,\AA$) occupy most of the probed $\xi_{\rm ion}$ parameter space, though with a median of $\log (\xi_{\rm ion} / {\rm Hz\,erg^{-1}})= 25.55$, higher than the canonical values, $\log (\xi_{\rm ion} / {\rm Hz\,erg^{-1}})= 25.1-25.3$ \citep{Robertson13}. The large scatter in $\xi_{\rm ion}$ at $D_{\rm Ly\alpha} > 50\,\AA$ could indicate bursty star formation, particularly since $\xi_{\rm ion}$ traces massive O-stars, and may reflect typical younger stellar populations that are yet to ionize the neutral \hi\ gas in their local surroundings. By contrast, galaxies that are strong LAEs tend to exhibit high ionizing photon production efficiencies, occupying the region with $\log (\xi_{\rm ion} / {\rm Hz\,erg^{-1}})> 25.5$, higher than the canonical value. Since these are predominantly found at $z<8$, we argue that it is mainly this particular subset of galaxies with high $\xi_{\rm ion}$ and late in the reionization epoch that provides the strongest contribution to this last phase transition.

\section{Discussion} \label{sec:disc}

To place our results into context of the large-scale efforts to quantify the galaxy mass assembly and reionization timeline, we describe in this section a physical scenario connecting the observed prevalence and prominence of \lya emission and DLA absorption from abundant, pristine \hi\ gas in galaxies through redshift and as a function of their intrinsic physical properties. We further quantify the effect on \dlya\ from the neutral hydrogen fraction based on more sophisticated IGM simulations and estimate the effect of strong galaxy DLAs on photometric redshifts measurements at $z>8$, which appear to be systematically overestimated.  

\subsection{\hi\ gas assembly and reionization across cosmic time} 

In this work, we quantified the strength of DLA absorption (\lya emission) via the positive (negative) strength of \dlya. We found that galaxies with high \dlya, indicating substantial \hi\ gas column densities in the ISM or CGM of galaxies, are ubiquitously observed at $z>9$ and in galaxies spanning the entire dynamical range of physical properties probed here. The fact that there appears to be no correlation between the amount of local \hi\ gas (or the lack thereof) and the ionizing output of the galaxies themselves may suggest that DLAs trace particularly young galaxies, at all redshifts, that are yet to ionize the surrounding \hi\ gas or convert it into molecules and eventually stars. For galaxies with $D_{\rm Ly\alpha} > 50\,\AA$, equivalent to $N_{\rm HI} > 10^{21}\,$cm$^{-2}$, the absorbing \hi\ gas is completely self-shielding of ionizing photons, even with substantial clumping with density variations down $\Delta N_{\rm HI} \approx 10^{-3}$ supporting this scenario. 

On the contrary, strong LAEs are only observed at $z\lesssim 8$. This indicates that only after $z\approx 8-9$, on average, is the surrounding large-scale IGM sufficiently ionized for a prominent fraction of \lya\ photons to escape. Indeed, the observed redshift evolution of \dlya\ is in good qualitative agreement with the evolution of the \lya\ escape fraction, as compiled by \citet{Saxena23b,Hayes23}.
Consequently, this favors more late and rapid reionization models \citep[e.g.,][]{Naidu20}, over a smooth, early transition \citep[e.g.,][]{Finkelstein19}. A similar rapid evolution seems to be suggested by damped \lya\ wings in quasar spectra \citep{Greig17,Greig19, Davies18,Banados18,Yang20,Wang21,Ocvirk21,Durovcikova20,Durovcikova24} and \lya\ emission statistics of galaxies \citep{Ouchi10,Sobacchi15,Mason18,Mason19}, indicating a near-fully neutral IGM at $z\approx 8$.

The strong LAEs are predominantly associated with UV faint ($M_{\rm UV} > -19$\,mag) galaxies with high ionizing photon production efficiencies, $\log (\xi_{\rm ion} / {\rm Hz\,erg^{-1}}) > 25.5$. This could either indicate that these particular systems were the most efficient at ionizing their immediate surroundings, or that viewing angles play a larger effect at low ($z<8$) redshifts compared to their higher redshift counterparts. In this scenario, galaxies at $z>8$ are deeply embedded into dense, neutral gas overdensities with large covering fraction, showing strong absorption independent of the viewing angle. At lower redshifts, low-density $N_{\rm HI}$ channels are carved by ionizing photons and we thus observe a broader span in \dlya\ at $z<8$. This scenario would be 
commensurate with the non-dependence of the strength of LAEs with the $O_{32}$ ionization parameter, [\oiii]+H$\beta$ EW, the $\beta_{\rm UV}$ spectral slope and the inferred escape fraction of LyC photons. In fact, strongly directionally-dependent escape of ionizing radiation has been proposed by simulations \citep{Cen15,Rosdahl22,Yeh23}. Since only $\approx 10-20\%$ of the galaxies at $z<8$ have strong LAEs, we argue that it is either this particular subset of faint, low mass galaxies that drive reionization or that ionizing photons from galaxies at this epoch will only be able to escape through the $\lesssim 20\%$ low-density sightlines.

\subsection{Insights into the \dlya\ parameter from simulations}

So far, the IGM regime of \dlya\ has been investigated assuming a homogeneously ionized IGM with an average value of the ionized fraction $x_{\rm HI}$ everywhere. It is therefore interesting to investigate how the \dlya\ parameter evolves in simulations of patchy reionization, to better understand the allowed range of  \dlya\ and what some limitations may be. 

Attenuation by the intergalactic medium is expected to play a significant role in regulating visibility of \lya\ emission until the Universe was fully reionized \citep{McQuinn07}, and will impact the \dlya\ parameter to some extent over all of the regimes identified in Fig. \ref{fig:dlyaz}. A full comparison of observed and simulated \dlya\ across the whole parameter space would therefore require detailed modelling of the redshift evolution of ionized bubbles, as well as the detailed shape and of the Ly$\alpha$ emission profile \citep[e.g.,][]{Hayes23}. We instead focus here on identifying the allowed range of \dlya\ for which attenuation by the IGM dominates the measurement. We measure \dlya\ from IGM transmission spectra normalised by the continuum, i.e., with no contribution from \lya\ emission.

\begin{figure}[!t]
    \centering
    \includegraphics[width=9cm]{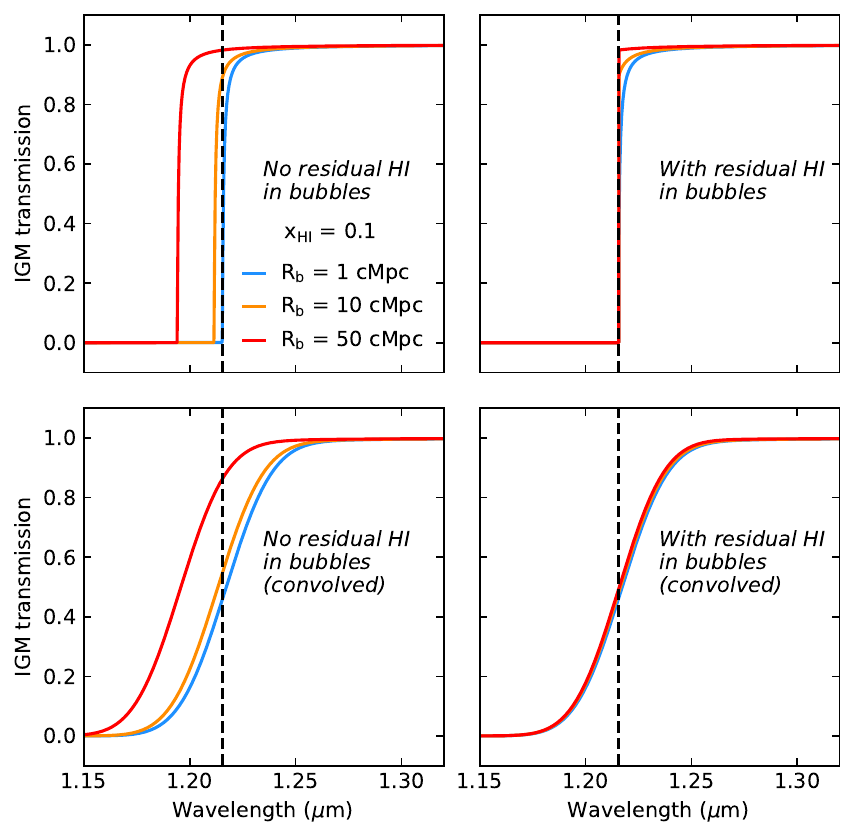}
    \caption{The impact of considering residual neutral hydrogen on IGM transmission curves for models with ionized bubble sizes of 1 cMpc (blue), 10 cMpc (orange) and 50 cMpc (red) and a volume-averaged IGM neutral fraction $x_{\rm HI} = 0.1$ at redshift 9. The vertical black dashed line marks the observed Ly$\alpha$ wavelength at this redshift. The left column shows IGM damping wings calculated using the \citet{MiraldaEscude98} model, assuming that the bubbles are completely ionized. The right column shows the effect of accounting for a small amount of residual neutral gas in the bubbles, which can completely saturate the absorption in the Ly$\alpha$ forest. The top panel shows the models before convolving with the NIRSpec PRISM instrumental profile and the bottom panel shows the models after convolution.}
    \label{fig:lyatranssims}
\end{figure}

First, we show how the \lya\ transmission curves change based on the residual fraction of \hi\ inside an ionized bubble. As discussed in several works, even a small amount of neutral gas can completely saturate absorption of the \lya\ forest \citep{Mesinger07,Mason18,Keating23a}. The difference in accounting for this residual neutral gas is shown in Fig.~\ref{fig:lyatranssims}, which shows IGM transmission curves calculated using the \citet{MiraldaEscude98} analytic model. In the top left panel, the bubbles are assumed to be completely ionized, allowing for transmission on the blue side of \lya. When this neutral gas is accounted for, a sharp cutoff in IGM transmission blueward of the \lya\ emission is observed. The bottom panel shows the same transmission profiles, but now convolved with the NIRSpec PRISM instrumental profile. It is difficult to statistically disentangle the difference between the models with different bubble sizes once the residual \hi\ is included.

\begin{figure}[!t]
    \centering
    \includegraphics[width=9.2cm]{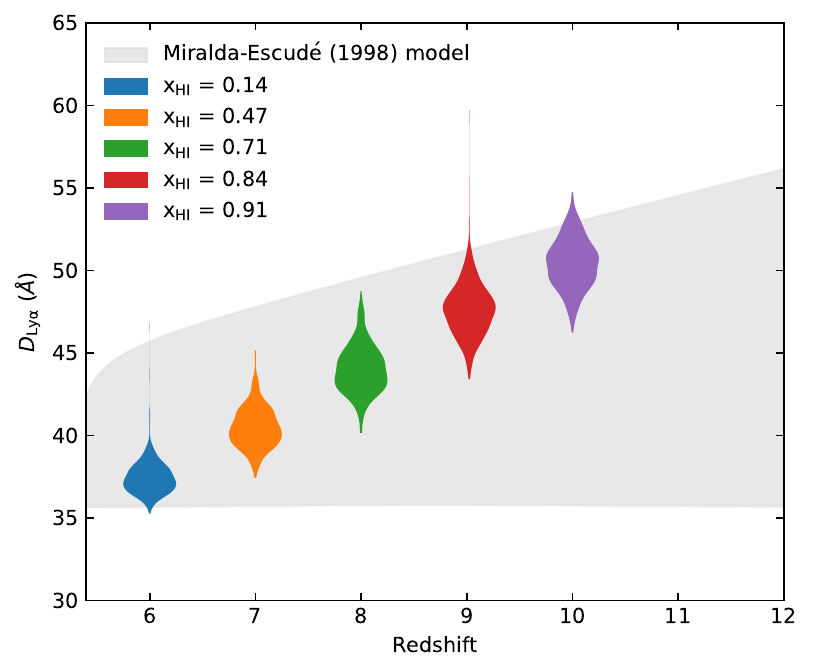}
    \caption{Redshift evolution of the damping parameter for continuum-normalised spectra dominated by IGM absorption. The grey shaded region shows the range of damping parameters calculated from a step function at the observed \lya\ wavelength (which sets the lower limit) and the IGM damping wing calculated using the \citet{MiraldaEscude98} model in a completely neutral Universe (which sets the upper limit). The coloured violin plots represent the distribution of damping parameters measured from IGM damping wings generated from the Sherwood-Relics simulation of an inhomogeneously reionized IGM as presented in \citet{Keating23a}. The volume-averaged \hi\ fraction $x_{\rm HI}$ for each simulation snapshot is indicated in the legend.}
    \label{fig:dlyaigmsims}
\end{figure}
 
Based on these results, we estimate an allowed range of IGM-dominated \dlya\ parameters, shown in Fig.~\ref{fig:dlyaigmsims}. The lower value of \dlya\ is given by a step function, where there is no transmission blueward of \lya\ and complete transmission redward of \lya\. This mimics the limit of large bubbles containing a small amount of residual \hi. The upper value of \dlya\ is calculated from the \citet{MiraldaEscude98} analytic model assuming the Universe is completely neutral and assuming a bubble size $R_{\rm b} = 0$ cMpc. We see that the maximum value of \dlya\ grows with redshift, due to the evolution of the mean density of the Universe \citep{Keating23b}.

For comparison, we also measure IGM-dominated \dlya\ parameters directly from damping wings constructed from a simulation of inhomogeneous reionization. We use the spectra presented and described in \citet{Keating23a}, which were generated from sightlines through a simulation from the Sherwood-Relics simulation suite \citep{Puchwein23}. We analyse results from five snapshots from this simulation at $z=6-10$. The distributions of \dlya\ parameters we measure are shown as the violin plots in Fig.~\ref{fig:dlyaigmsims}. We find that the range of allowed IGM \dlya\ parameters is well captured by the grey shaded region we defined. At $z=6$, some sightlines have smaller \dlya\ parameters as the \lya\ forest starts to show transmission at this redshift. At $z=10$, some sightlines have larger \dlya\ parameters due to effects not captured in the analytic model for IGM damping wings, such as dense clumps of gas close to the host halo or the effect of infalling material. In general, however, the range of \dlya\ parameters for which the reionization of the IGM plays a dominant role are in good agreement with the range of $D_{\rm Ly\alpha} = 35$--$50\,\AA$ as defined in Section \ref{ssec:dlya}. However, we note that the IGM will also play an important role in regulating the fraction of flux transmitted by galaxies showing evidence of \lya\ emission, and hence falling into the regime of smaller values of \dlya.

\subsection{The effect of \lya\ damping wings on high-$z$ photometric redshift estimates}

Some of the first efforts to photometrically identify high-redshift galaxy candidates seemed to generally overpredict their redshifts when compared to the subsequent spectroscopic redshift determination of the sources at $z>8$ based on strong nebular emission lines \citep[e.g.,][]{Finkelstein23b,Fujimoto23_ceers,Hainline23,Steinhardt23,Serjeant23}. In \citet{Heintz23_DLA}, we proposed that this may due to the strong damping wings observed in the large fraction fraction of galaxies at these redshifts, suppressing the derived flux density in the blue-most photometric bands and thereby mimicking a slightly larger \lya ``break redshift'' \citep[e.g.,][]{ArrabalHaro23}.

\begin{figure}[!t]
    \centering
    \includegraphics[width=9.2cm]{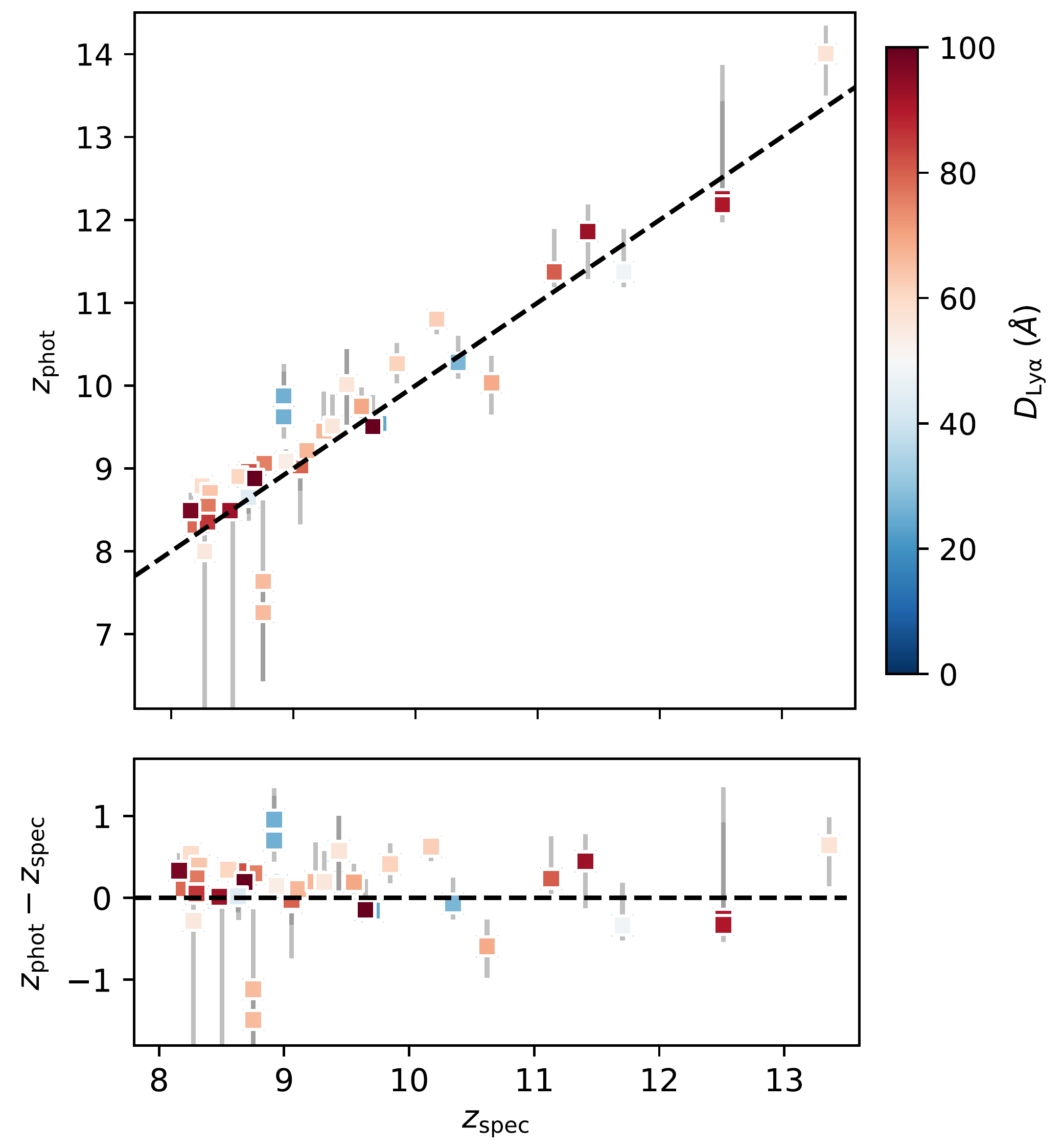}
    \caption{Photometric redshift, $z_{\rm phot}$, estimates relative to the spectroscopic redshifts, $z_{\rm spec}$, for the \jw-PRIMAL sample at $z>8$. The sources are color-coded as a function of \dlya, as indicated by the colorbar. The photometric redshifts systematically overpredicts the actual spectroscopically derived redshifts with median $z_{\rm phot}-z_{\rm spec} = 0.11$ ($z_{\rm phot}-z_{\rm spec} = 0.15$ for $D_{\rm Ly\alpha} > 50\,\AA$). }
    \label{fig:zphotzspec}
\end{figure}

To test this, we show the derived photometric redshifts for the subset of sources where these are available in the DJA \citep{Valentino23}\footnote{Available at \url{https://dawn-cph.github.io/dja/imaging/v7/}} relative to the spectroscopically derived redshifts in Fig.~\ref{fig:zphotzspec}. We find that the photometric redshifts are systematically overestimated for galaxies at $z>8$, with median $z_{\rm phot}-z_{\rm spec} = 0.11$. Considering only the galaxies with strong DLAs ($D_{\rm Ly\alpha} > 50\,\AA$) the discrepancy become slightly larger, with median $z_{\rm phot}-z_{\rm spec} = 0.15$. To optimize future photometric redshift estimates, the effects of strong \lya\ damping wings needs to be incorporated into spectral energy distribution fitting algorithms like {\sc Eazy} \citep{Brammer_eazy}. 




\section{Summary \& Future Outlook} \label{sec:conc}

In this work, we have compiled and systematically characterized 494 sources observed with \jw/NIRSpec during the reionization epoch, at $z=5.5$--13.4, introducing the \jw-PRIMAL legacy survey. The main goal was to statistically quantify the presence and redshift evolution of strong damped \lya\ absorption in galaxies, signifying abundant local \hi\ gas reservoirs. To disentangle these systems from sources with damping parameters probing mainly a largely neutral IGM, or residing in extended ionized bubbles, we defined a new simple diagnostic, the \lya\ damping parameter, \dlya. This allowed us to statistically chart the \hi\ gas assembly history, the onset and prevalence of strong \lya-emitting galaxies and to explore the main underlying physical properties driving the emission and escape of ionizing and \lya photons. 

We found that the majority ($\approx 65-90\%$) of galaxies at $z>8$ are dominated by large \hi\ gas reservoirs in their local surroundings (ISM or CGM) with galaxy-integrated \hi\ column densities, $N_{\rm HI}>10^{21}\,$cm$^{-2}$. These strong galaxy DLAs appear to exist throughout the reionization epoch, even though the fraction was observed to decrease to $\approx 60\%$ by $z=6$ of the overall galaxy population. Similar \hi\ column densities have been theoretically predicted from cosmological simulations \citep[e.g.,][]{Finlator18,DOdorico18,Pallottini22}. Further, we found that these strong DLA galaxies represent a large variety of intrinsic physical properties in terms of their UV luminosity, gas-phase metallicity $12+\log$(O/H), UV spectral slope $\beta_{\rm UV}$, nebular [\oiii]+H$\beta$ line EWs, $O_{32}$ ionization parameters, and ionizing photon production efficiency, $\xi_{\rm ion}$. We thus surmised that strong galaxy DLAs are a sign of early galaxies in the process of formation, actively accreting and assembling large amount of primordial \hi\ gas, that are yet to be ionized or processed into molecular gas and stars.  

From the measured redshift evolution of \dlya\ we were able to quantify the onset and increasing strength of \lya, enabling us to decode the reionization history directly. We found that at $z\approx 8$--10 galaxy spectra mainly probe a substantially neutral CGM, whereas evidence for low \lya\ escape or significant ionized bubbles around these galaxies were only observed at later cosmic times ($z<8$). The fraction of LAEs was estimated to be $\approx 20\%$ at $z=6$. 
The strongest LAEs at $z>7$ have been found to reside in particular overdense or ionized regions \citep{Saxena23,Witstok23,Witten24}. From theoretical considerations and cosmological simulations \citep{Pallottini22}, we expect most of the galaxies studied here to be formed within the first 200--300\,Myr of cosmic time at $z\approx 18$--14 and rapidly build up their stellar mass and on-going star formation via substantial \hi\ gas accretion and assembly. Over the next 100--300\,Myr, down to $z=8-100$, 

The emergence of these sources showing strong ``roll-overs'' near the \lya\ edge in the first era of \jw/NIRSpec Prism spectra have important practical implications for future studies and observations: First, they complicate statistical inferences of the reionization history by tracing the increasing neutral fraction of the IGM from galaxy damping wing measurements \citep{Keating23a}. Here, we provided the first step forward to statistically model the galaxy DLA population which will eventually be possible to remove to uncover the underlying IGM damping effects. Second, the first efforts in photometrically identifying high-redshift galaxy candidates seemed to show a systematic positive offset in $z_{\rm phot}-z_{\rm spec}$ towards higher redshifts \citep{Finkelstein23b,Fujimoto23_ceers,Hainline23,Steinhardt23,Serjeant23}. This may naturally be explained by the increasing fraction of galaxies with stronger \lya\ damping wings at higher redshifts. This we quantified here and estimated a mean of 0.2\,dex overestimations of the photometric redshifts of galaxies at $z>8$. This effect would need to be incorporated into photo-$z$ codes to optimize the accuracy on the $z_{\rm phot}$ priors for the earliest galaxy population. 

The next natural avenues to explore the early galaxy DLAs in more depth are to examine the effect of clustering, their correlations with physical sizes and SFR surface densities, and the \hi\ gas content; a potential cause of the observed offset in the otherwise fundamental-metallicity relation (FMR). Similar to how strong LAEs are typically found to reside in high-density groups of galaxies, galaxy DLAs may trace overdense regions via their early assembly of \hi. Moreover, the subset of galaxies showing strong DLAs integrated over their relative small UV sizes indicate exceptional dense gas surface densities \citep{Kennicutt12}. This, in combination with the inferred SFRs, may be key to understand the intense star formation and bright UV luminosities of early $z>9$ galaxies \citep[e.g.,][]{Finkelstein23,Franco23,Harikane23,Adams23,Casey23,Bouwens23,Chemerynska23,Mason23,McLeod24}. Finally, the discovered offset from the FMR towards lower metallicities of galaxies at $z=7$--10 have been interpreted as evidence for excessive pristine \hi\ gas inflow \citep{Heintz23_FMR}, which have later been confirmed with larger galaxy samples, though the exact redshift for this transition is still debated \citep[e.g.,][]{Curti23b,Nakajima23}. This scenario will be possible to test directly by correlating the \hi\ column densities of galaxies to their offset from the FMR, moving the 3D plane of galaxy properties to 4 dimensions. Obtaining higher-resolution spectroscopic data of these sources will further help quantifying the \hi\ covering fraction and disentangle the effects on the \lya\ transmission curve from weak \lya\ emission to extended ionized bubbles and expand the probed physical properties to include the electron densities and temperatures of the gas.       

While previous efforts with ALMA to quantify the \hi\ gas mass content using far-infrared tracers such as [\cii]$-158\mu$m and [\oi]$-63,145\mu$m have been successfully applied to galaxies in the epoch of reionization \citep{Heintz21,Heintz22,Vizgan22,Liang24,Wilson23}, the results presented here provide the most direct and statistical census of the abundance and assembly of the primordial \hi\ gas in early Universe. This naturally paves the way and provide an important benchmark for next-generation radio facilities, targeting \hi\ of the first galaxies through the 21-cm transition, such as the square kilometre array \citep[SKA;][]{Dewdney09}.

\section*{Acknowledgements}

We would like to thank Peter Jakobsen for his vision and heroic endeavour in optimally designing the \jw/NIRSpec instrument and some of its first on-sky observations and for enlightening discussions about the intricacies of the NIRSpec data. Further, we would like to thank John Chisholm for helpful clarifications and discussions related to the escape fraction of ionizing photons and Aayush Saxena for enlightening conversations on the escape and absorption of Lyman-$\alpha$ photons.

This work has received funding from the Swiss State Secretariat for Education, Research and Innovation (SERI) under contract number MB22.00072.
The Cosmic Dawn Center (DAWN) is funded by the Danish National Research Foundation under grant DNRF140.
The data products presented herein were retrieved from the DAWN \jw Archive (DJA). DJA is an initiative of the Cosmic Dawn Center, which is funded by the Danish National Research Foundation under grant DNRF140.
P.D. acknowledge support from the NWO grant 016.VIDI.189.162 (``ODIN") and warmly thanks the European Commission's and University of Groningen's CO-FUND Rosalind Franklin program. 
Support from the ERC Advanced Grant INTERSTELLAR H2020/740120 is kindly acknowledged (A.F.).
S.G. acknowledges financial support from the Villum Young Investigator grants 37440 and 13160 and the Cosmic Dawn Center.
M.K. was supported by the ANID BASAL project FB210003. 
G.E.M. acknowledges financial support from the Villum Young Investigator grants 37440 and 13160 and the Cosmic Dawn Center. 
J.W. acknowledges support from the Science and Technology Facilities Council (STFC), by the ERC through Advanced Grant 695671 ``QUENCH'', by the UKRI Frontier Research grant RISEandFALL. Support for this work was provided by NASA through the NASA Hubble Fellowship grant HST-HF2-51515.001-A awarded by the Space Telescope Science Institute, which is operated by the Association of Universities for Research in Astronomy, Incorporated, under NASA contract NAS5-26555. 
F.C. acknowledges support from a UKRI Frontier Research Guarantee Grant (PI Cullen; grant reference EP/X021025/1).
J.H.W. acknowledges support by NSF grant AST-2108020 and NASA grants 80NSSC20K0520 and 80NSSC21K1053.
NRT acknowledges support through STFC consolidated grant ST/W000857/1.  M.J.H. is supported by the Swedish Research Council, Vetenskapsr$\aa$det, and is fellow of the Knut \& Alice Wallenberg foundation. 

This work is based in part on observations made with the NASA/ESA/CSA James Webb Space Telescope. The data were obtained from the Mikulski Archive for Space Telescopes (MAST) at the Space Telescope Science Institute, which is operated by the Association of Universities for Research in Astronomy, Inc., under NASA contract NAS 5-03127 for \jw.

\section*{Data availability statement} 

All the data presented in this work are made publicly available through DJA: \url{https://dawn-cph.github.io/dja/}, and are made easy to inspect and query in a dedicated webpage: \url{https://s3.amazonaws.com/msaexp-nirspec/extractions/nirspec_graded_v2.html}. The spectroscopic data have been processed using the custom-made reduction pipeline \textsc{MsaExp}, publicly available here: \url{https://github.com/gbrammer/msaexp}. All the data and catalogs related to \jw-PRIMAL are also available on a dedicated webpage: \url{https://github.com/keheintz/jwst-primal}.  

Software: This work made use of and acknowledge the following software: {\tt NumPy} \citep{Numpy}, {\tt Matplotlib} \citep{matplotlib}, {\tt LMfit} \citep{lmfit}, {\tt SciPy} \citep{scipy}, \texttt{grizli} \citep{Brammer_grizli}, {\tt Astrodrizzle} \citep{AstroDrizzle}, and {\tt MsaExp} \citep[v0.3;][]{Brammer_msaexp}. 

\bibliographystyle{aa}
\bibliography{ref.bib}











\end{document}